\documentclass[preprint,12pt]{elsarticle}

\usepackage{amssymb}
\usepackage{amsthm}
\usepackage{xcolor}
\usepackage{graphicx}
\usepackage{lineno,hyperref}
\modulolinenumbers[5]

\usepackage{inputenc}
\usepackage{amsmath}
\usepackage{graphics}
\usepackage{epsfig}
\usepackage{graphicx}
\usepackage{latexsym}
\usepackage{graphics}
\usepackage{epsfig}
\usepackage{amsmath,amssymb,amsfonts,theorem,color,bm}
\usepackage{multirow}
\usepackage{stfloats} 
\usepackage{float}
\usepackage{subfig}
\usepackage{soul}
\usepackage{caption}
\usepackage{hyperref}
\hypersetup{colorlinks,allcolors=black}

\newcommand{\bI}{{\bf I}}
\newcommand\bc{\boldsymbol c}

\newcommand\bu{\boldsymbol u}

\newcommand\bv{\boldsymbol v}
\newcommand\bV{\boldsymbol V}

\newcommand\bW{\boldsymbol W}

\newcommand\bo{\boldsymbol 0}

\newcommand\bS{\boldsymbol S}
\newcommand\bT{\boldsymbol T}

\newcommand\bPhi{\boldsymbol{\Phi}}
\newcommand\bSigma{\boldsymbol{\Sigma}}

\newcommand\bR{\boldsymbol{R}}

\numberwithin{equation}{section}

\newcommand{\beqn}{\begin{equation}}
\newcommand{\eeqn}{\end{equation}}
\newcommand{\beqnarr}{\begin{eqnarray}}
\newcommand{\eeqnarr}{\end{eqnarray}}
\newcommand{\baling}{\begin{alignat}{1}}
\newcommand{\ealing}{\end{alignat}}

\usepackage{xcolor,colortbl}
\definecolor{Gray}{gray}{0.75}
\newcolumntype{a}{>{\columncolor{Gray}}c}

\journal{Computer Physics Communications}

\begin{document}

\begin{frontmatter}



\title{ModelFLOWs-app: data-driven post-processing and reduced order modelling tools}


\author[UPM]{Ashton Hetherington}

\author[UPM]{Adrián Corrochano}

\author[UPM]{Rodrigo Abadía-Heredia}

\author[UPM]{Eneko Lazpita}

\author[ULB,BRITE]{Eva Muñoz}

\author[UPM]{Paula Díaz}

\author[UPM]{Egoitz Maiora}

\author[UPM]{Manuel López-Martín}

\author[UPM]{Soledad Le Clainche\footnote{Correspondence to: soledad.leclainche@upm.es}}

\affiliation[UPM]{organization={ETSI Aeronáutica y del Espacio, Universidad Politécnica de Madrid},
            addressline={Plaza Cardenal Cisneros, 3}, 
            city={Madrid},
            postcode={28040}, 
            country={Spain}}

\affiliation[ULB]{organization={Aero-Thermo-Mechanics Department},
            addressline={Université Libre de Bruxelles}, 
            city={Brussels},
            postcode={1000}, 
            country={Belgium}}

\affiliation[BRITE]{organization={Brussels Institute for Thermal-fluid systems and clean Energy (BRITE)},
            addressline={Université Libre de Bruxelles and Vrije Universiteit Brussel}, 
            city={Brussels},
            postcode={1000}, 
            country={Belgium}}
            
\begin{abstract}
This article presents an innovative open-source software named ModelFLOWs-app\protect\footnote{The website of the software is available at \href{https://modelflows.github.io/modelflowsapp/}{https://modelflows.github.io/modelflowsapp/}}, written in Python, which has been created and tested to generate precise and robust hybrid reduced order models (ROMs) fully data-driven. By integrating modal decomposition and deep learning methods in diverse ways, the software uncovers the fundamental patterns in dynamic systems. This acquired knowledge is then employed to enrich the comprehension of the underlying physics, reconstruct databases from limited measurements, and forecast the progression of system dynamics. These hybrid models combine experimental and numerical database, and serve as accurate alternatives to numerical simulations, effectively diminishing computational expenses, and also as tools for optimization and control.  The ModelFLOWs-app software has demonstrated in a wide range of applications its great capability to develop reliable data-driven hybrid ROMs, highlighting its potential in understanding complex non-linear dynamical systems and offering valuable insights into various applications. This article presents the mathematical background, review some examples of applications and introduces a short tutorial of ModelFLOWs-app.
\end{abstract}



\begin{keyword}
open-source software \sep reduced order models \sep data analysis \sep deep-learning \sep patterns identification \sep data-driven methods
\end{keyword}

\end{frontmatter}



\section{Introduction\label{sec:introduction}}

The availability of high-quality data has sparked a revolution in machine learning and reduced order modeling. Data-driven equation-free models offer a promising approach to understanding complex non-linear dynamical systems, even without prior knowledge of the governing equations. Machine learning tools allow us to extract knowledge directly from the data, rather than relying solely on theoretical principles. This shift encourages the use of data to uncover new hypothesis and models. While these techniques present challenges, they also provide significant opportunities for advancing our understanding of such systems, with applications in various industries including aerospace, automotive, construction, pharmaceuticals, chemicals, manufacturing, and more.

There are two primary approaches to data-driven modeling. The first approach involves data forecasting models, which focus on predicting future data using machine learning techniques like deep neural networks. These models do not incorporate explicit physical insights into their construction. The second approach is represented by reduced order models (ROMs) with physical insights. These hybrid models incorporate physical understanding and utilize pattern identification techniques such as proper orthogonal decomposition \cite{Sirovich87} or dynamic mode decomposition \cite{Schmid10} to extract relevant spatio-temporal information from the data.

By employing hybrid data-driven ROMs in non-linear dynamical systems, several advantages emerge over relying solely on deep neural networks. These ROMs enable the identification of key instabilities and mechanisms within the studied database by capturing important information about the underlying physics. Furthermore, they facilitate the development of powerful tools for optimization and control. Having a deeper physical understanding of the problem allows for enhanced system phase prediction, adoption of more controlled and robust strategies, reduction in computational costs for numerical simulations, and streamlined information collection in experiments.

This article introduces the methodology and some relevant resutls obtained with a totally novel software,  named as ModelFLOWs-app, which has been tested suitable to develop accurate and robust fully data-driven hybrid ROMs. Modal decomposition and deep learning strategies are combined in several ways to reveal the main  patterns in dynamical systems, and uses this information to understand the physics of the problem under study, to reconstruct databases from sparse measurements or to predict the evolution of the system dynamics, providing accurate alternatives to numerical simulations, reducing in this way related computational cost. 

The origin of this tool was for the analysis of complex flows, where ModelFLOWs-app has been tested in a wide range of applications, for instace, (i) for temporal forecasting in reactive flows \cite{paperAdricombustion}, wind turbines \cite{LeClaincheFerrer18}, compressible flows with buffeting \cite{KouLeClaincheZhangPoF18}, etc., (ii) for identification of flow instabilitites and patterns in wall bounded turbulent flows \cite{LeClaincheetalJFM20,LeClaincheetalJFM22}, synthetic jets \cite{LeClainche19}, urban flows \cite{Lazpita}, etc., (iii) for data reconstruction in experiments modelling turbulent or complex flows \cite{PaperAvelazquezJM,PaperAvelazquezAIAApaper,paperPaula}, etc. Additionally,  due to the robustness and exceptional properties presented by the models and algorithms composing ModelFLOWs-app, the software has been tested and extended to a wide range of industrial applications solving problems in complex non-linear dynamical systems. Some examples of the multiple applications of this tool include,  patterns identification and reconstruction in medical imaging \cite{NoureletalHODMD,NoureletalNMR}, wind velocity predictions with LiDAR experimental measurements \cite{LeClaincheLorenteVegaEnergies18}, identification of cross-flow instabilitites \cite{LeClaincheFerrer18}, prediction of flutter instability in flight test experiments \cite{LeClaincheetalJAircraft18,paperCarlosFlightTest},  to name a few. We invite the readers to review carefully the applications presented in this article and to complement their knowledge with the tutorials and videos presented in ModelFLOWs-app website \cite{ModelFLOWsappWeb}.

The article is organized as follows. A general description of the methodological framework behind ModelFLOWs-app is introduced in \S\ref{sec:methodology}, and the methodology connected to the two big modules forming this software is shown in \S\ref{sec:module1} and \S\ref{sec:module2}. A review of some of the main results of ModelFLOWs-app is presented in \S\ref{sec:results}. Finally, \S\ref{sec:conclusions} explain the main conclusions. 

\section{Methodology\label{sec:methodology}}
\begin{figure}[!htbp]
\centering
\includegraphics[trim=60 0 60 50, clip, width=0.85\textwidth]{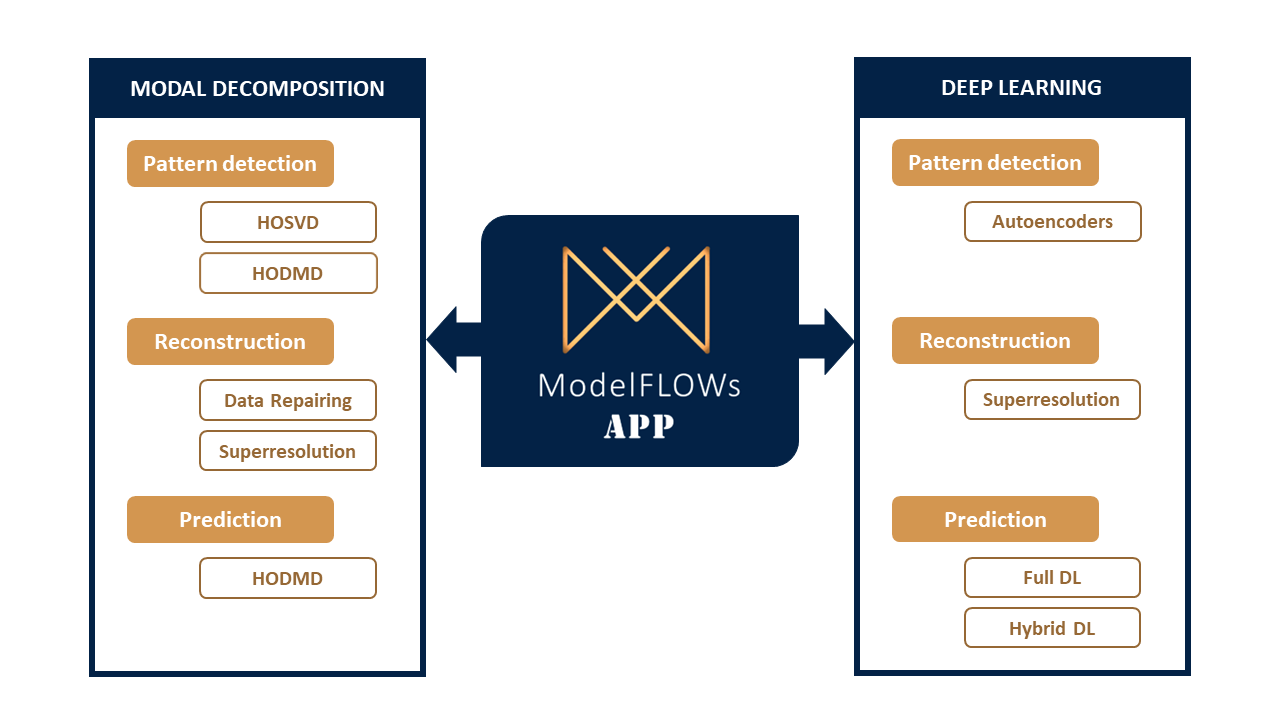}
\caption{General distribution of ModelFLOWs-app software.}
\label{fig:app}
\end{figure}

ModelFLOWs-app methodological framework is formed by two big modules: Module 1 uses modal decomposition methods, and Module 2 is formed by hybrid machine learning tools, which combine modal decomposition with deep learning architectures. Each module solves three different applications: (1) patterns identification, suitable to study the physics behind the data analysed; (2) data reconstruction, capable to reconstruct two- or three- dimensional databases from a set of selected points, using data from sensors, or repairing missing data; (3) data forecasting, which builds reduced order models (ROMs) to predict the spatio-temporal evolution of the signal analysed.  Figure \ref{fig:app} shows an sketch with the general organization of the software.

This section introduces the mathematical principles behind each Module, and also provides some examples of application. A detailed guide on how to treat and analyse the selected databases to exploit the properties of this software is presented in the ModelFLOWs-app website \cite{ModelFLOWsappWeb}. 
The software is presented in two different modalities. The first one is the application-web, which is mounted on an interface over {\it streamlit~\cite{streamlit}.} This modality is suitable to understand what are the possibilities of ModelFLOWs-app, what are the setting parameters to run each application, what are the input and output files and to understand the capabilities of the software. 
The second one is formed by the main code, written in  Python \cite{python}, which calls the modules and applications. This modality is suitable to be used in complex problems and big databases. The codes are open to the general public, so it can be locally modified and adapted to the user needs.

\subsection{Data organization}

ModelFLOWs-app is a fully data-driven software, so the user should first provide the desired databases to analyze and next select the application to solve. 

In matrix form, a database  is formed by a
 set of $K$ snapshots $\bv_k=\bv(t_k)$, where $t_k$ is the time measured at instant $k$, that for convenience,
 are collected in the following
 {\it snapshot matrix}
  \beqn
  \bV_1^K = [\bv_{1},\bv_{2},\ldots,\bv_{k},\bv_{k+1},\ldots,\bv_{K-1},\bv_{K}].\label{ab0}
  \eeqn

For some applications, it is more convenient to re-organize the database into tensor form, called as {\it snapshot tensor}. 
In that case, the various components that constitute the database are separated and re-organized into different tensor components, and similarly, the different spatial coordinates are also separated too. Generally, the database components are formed by velocity components (especially in fluid dynamics applications), although it is possible to consider any type of variable depending on the application solved. For instance, in combustion databases the components are formed by the different species \cite{paperAdricombustion}, to identify flutter instability in flight test, the components are formed by the signal given by an array of accelerometers \cite{paperCarlosFlightTest}, in atmospheric boundary layer flows, in addition to the velocity components, the pressure is included \cite{paperPaula}, etcetera. 

\begin{figure}[h!]
	\begin{center}
    \includegraphics[trim=150 200 150 180, clip, width=\textwidth]{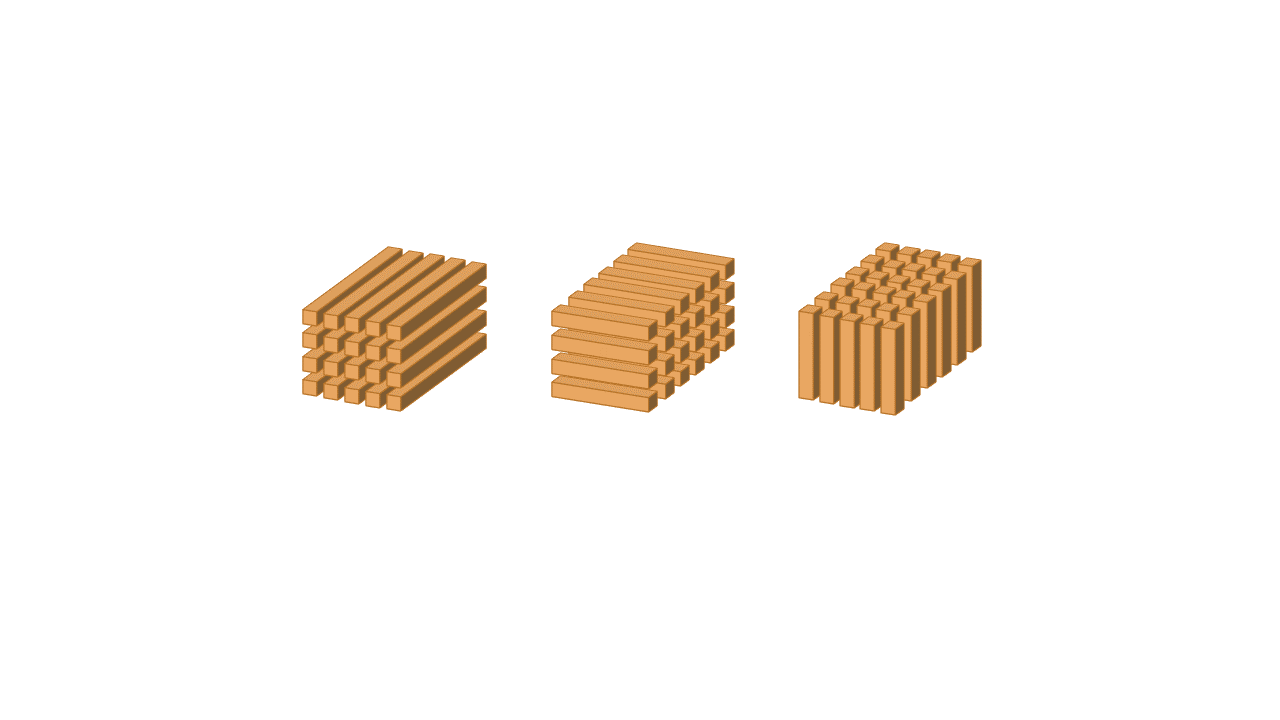}
\vskip-0.1cm
\end{center}
\vskip-0.5cm
	\caption{The fibers of a third order tensor. \label{fig:tensorFiber}}
\end{figure}

In the snapshot tensor, the matrix snapshots fit a multidimensional array, which depends on more than two indexes. The {\it fibers} of the tensor are formed by the corresponding matrix columns and rows.  Fig.\ref{fig:tensorFiber} shows an example of a third order tensor, where the tensor fibers are identified.

Generally, the algorithms presented in ModelFLOWs-app consider fourth and fifth order tensors for the analysis of two- and three-dimensional databases, respectively. For instance, let us consider a two-dimensional database (plane) formed by two velocity components (although, as mentioned before, the type and number of components are dependent on the database studied),  streamwise and normal velocity components $v_x$ and $v_y$, of the
  in-plane velocity $\bv$
 in a Cartesian coordinate system with dimension $J_2\times J_3$, as
\beqn
\bv(x_{j_2},  y_{j_3},t_k)
\quad\text{for }j_2=1,\ldots, J_2, \quad j_3=1,\ldots, J_3, \quad k=1\ldots, K.\label{c2}
\eeqn

The snapshots can be re-organized in a fourth-order $J_1\times J_2\times J_3\times K$-tensor $\bV$,
 whose components $V_{j_1j_2j_3k}$ are defined as
\beqn
V_{1j_2j_3k}=v_x(x_{j_2},y_{j_3},t_k),\quad V_{2j_2j_3k}=v_y(x_{j_2},y_{j_3},t_k).\label{c4}
\eeqn

The indexes $j_1$, $j_2$, $j_3$ and $k$ labels the velocity components (in this particular case $j_1=1,2$, where $J_1=2$), the discrete values of the two spatial coordinates, and the values of time. 
Note that, although we present a particular case for simplicity, the snapshot tensor $V_{j_1j_2j_3k}$, can be used with different number of components $J_1$, satisfying the need of the problem understudy. 

For three-dimensional time-dependent databases, the database is organized in a fifth-order $J_1\times J_2\times J_3\times J_4\times K$-tensor $\bV$,
 whose components $V_{j_1j_2j_3j_4k}$ are defined as 
 \beqn
 \begin{split}
V_{1j_2j_3k}&=v_1(x_{j_2},y_{j_3},z_{j_4},t_k),\\ V_{2j_2j_3k}&=v_2(x_{j_2},y_{j_3},z_{j_4},t_k),\\
\cdots\\
V_{j_1j_2j_3k}&=v_2(x_{j_2},y_{j_3},z_{j_4},t_k),\\
\cdots\\
V_{J_1j_2j_3k}&=v_{J_1}(x_{j_2},y_{j_3},z_{j_4},t_k).\label{c43d}
\end{split}
\eeqn
$J_1$ is the number of selected components forming the database (i.e., three velocity components), the indexes $j_2$, $j_3$ and $j_4$ correspond to the discrete values of the three spatial coordinates, $x$, $y$ and $z$, which are streamwise, normal and spanwise components, and $k$ is the index representing the time instant.

Let us note that the simple connection between the  snapshot matrix, presented in eq. (\ref{ab0}), and the present snapshot tensor formulation. In the snapshot matrix, the tensor indexes $j_1$, $j_2$, $j_3$ (and $j_4$ for three-dimensional databases) are folded together into a single index $j$. So, $\bV_1^K\in J\times K$, where $J=J_1 \times J_2 \times J_3 (\times J_4)$. Hence, data can be easily transformed from matrix to tensor and tensor to matrix just by re-shaping the database, using the {\it reshape} function found in Numpy in Python\cite{python}, as it is done in ModelFLOWs-app.

\subsection{Relative root mean square error (RRMSE)}
 
The relative root mean square error (RRMSE) is computed to measure the quality of the ROMs developed in each one of the modules as
  \beqn
  RRMSE=\sqrt{\frac{\sum_{k=1}^K||\textbf{v}_k-\textbf{v}^{approx.}_k||^2}{\sum_{k=1}^K||\textbf{v}_k||^2}},\label{eq:rrmse}
  \eeqn
  where $||\cdot||$ is the usual Euclidean norm and vectors $\textbf{v}_k$ and $\textbf{v}^{approx.}_k$ correspond to the real and approximated solution.

\section{Module 1: modal decomposition\label{sec:module1}}

\subsection{Singular value decomposition (SVD) and proper orthogonal decomposition (POD)\label{sec:SVD}}

Lumley\cite{Lumley} introduced Proper Orthogonal Decomposition (POD) as a mathematical approach for extracting coherent structures from turbulent flows. The primary objective of POD is to decompose data into modes that optimize the mean square of a field variable under analysis. The classical method to calculate POD modal expansion  is based on the covariance of a state vector that changes over time, with the size of the state vector being based on the spatial degrees of freedom of the data. This method becomes extremely computationally expensive for large two-dimensional or three-dimensional problems. In such cases, a different technique, singular value decomposition (SVD) is used to obtain the POD modes, introduced by Sirovich \cite{Sirovich87}. 

In fluid dynamics, specifically in the study of turbulent flows, SVD is a widely used factorization technique. SVD captures the directions of a matrix which represent the vectors that can either shrink or grow. These directions are determined by the eigenvalues and eigenvectors of a rectangular matrix. In addition to fluid dynamics, SVD has found a wide range of applications, particularly in low-rank matrix approximations. This approach is beneficial because it reduces the size of the data being analyzed, removes noise and filters spatial redundancies \cite{LeClaincheetalJFM20}.

It is remarkable that the literature often uses the terms SVD and POD interchangeably, but SVD is one of the two possible techniques that can applied to obtain a POD decomposition.

SVD (or POD) decomposes a collection of spatio-temporal data $\bv(x,y,z,t)$ into a set of proper orthogonal spatial modes, also known as SVD or POD modes, represented by $\bPhi_n(x,y,z)$, which are weighted by the temporal coefficients $\bc_n(t)$, as
\begin{equation}
\bv(x,y,z,t)\simeq \sum_{n=1}^N \bc_n(t)\bPhi_n(x,y,z).\label{ab00pod}
\end{equation}

SVD algorithm factorizes the snapshot matrix $\bV_1^k$, eq. (\ref{ab0}), which is decomposed into the spatial orthogonal modes $\bW$, the SVD or POD modes, temporal modes $\bT$ and singular values $\bSigma$ as
\begin{equation}
\bV_1^{K}\simeq\bW\,\bSigma\,\bT^\top.\label{ab20}
\end{equation}
where $()^\top$ denotes the matrix transpose. The diagonal of matrix $\bSigma$ contains the singular values $\sigma_1,\cdots,\sigma_{K}$, $\bW^\top\bW = \bT^\top\bT=$ the $N\times N-$unit matrix, being $N$  the number of SVD modes retained. This parameter is also called {\it spatial complexity}, and will be referred in the following sections, when DMD-based methods will be introduced. It is remarkable the difference between $J$, the {\it spatial dimension}  of the database, and the spatial complexity $N$, where $N\leq min(J,K)$.

SVD modes are ranked in descending order based on their singular values. Typically, the modes with the highest singular values embody the system's general dynamics, representing coherent structures or patterns,  meanwhile, the modes with the smallest singular values may be omitted from the approximation, assuming a certain level of error. These modes could be related to noise, especially in the case of experimental databases, with spatial redundancies or, in fluid dynamics, they are generally connected (although not always) to small size flow scales.

The number of $N$ SVD modes to be retained in the approximation, to construct the expansion eq. (\ref{ab00pod}), can be determined using different criteria, as discussed in Ref. \cite{PCA} (in this context, the POD or SVD approaches are also referred to as \textit{Principal Component Analysis} - PCA). In the present context, the number of retained modes are estimated for a certain tolerance (tunable), $\varepsilon_{svd}$, which could be comparable to the level of noise (in the case of experimental results), could be connected to the size of the flow structures (in turbulent flows), et cetera, defined as 
\beqn
\sigma_{N+1}/\sigma_{1}\leq \varepsilon_{svd}.\label{eq:TOLsvd}
\eeqn
%


\subsection{High order singular value decomposition (HOSVD)\label{sec:HOSVD}}

Introduced by Tucker in 1966 \cite{Tucker66}, the HOSVD algorithm has gained popularity in recent years, particularly due to its implementation by de Lathauwer et al. \cite{DeLathawer,DeLathawer0}. This algorithm has proven to be effective in diverse fields such as aeronautic database generation \cite{LorenteVV08}, database compression \cite{LorenteVV10}, conceptual aeronautic design \cite{deLucasVV11}, and real-time control of automotive engines \cite{BenitoAVV11}.

HOSVD decomposes datbases organized in tensor form, where SVD is applied to each one of the fibers of the tensor. For instance,  HOSVD of the fifth order tensor defined in eq. (\ref{c43d}) is presented as
 \begin{equation}
 V_{j_1j_2j_3j_4k}\simeq\sum_{p_1=1}^{P_1}\sum_{p_2=1}^{P_2}\sum_{p_3=1}^{P_3}\sum_{p_4=1}^{P_4}\sum_{n=1}^{N} S_{p_1p_2p_3p_4n}
   W^{(1)}_{j_1p_1} W^{(2)}_{j_2p_2} W^{(3)}_{j_3p_3} W^{(4)}_{j_4p_4}T_{kn},    \label{c10}
 \end{equation}
where $\bS_{p_1p_2p_3p_4n}$ is the {\em  core tensor}, another fifth-order tensor, and the columns of the matrices
$\bW^{(1)}$, $\bW^{(2)}$, $\bW^{(3)}$, $\bW^{(4)}$ are $\bT$ are known as the {\it modes} of the decomposition. 

The first set of modes (i.e, the columns of the matrices $\bW^{(l)}$ for $l=$1,2,3 and 4) correspond to the number of components of the database and the spatial variables, so they are known the spatial HOSVD modes, while the columns of the matrix $\bT$ correspond to the time variable, these are the temporal HOSVD modes.

The singular values of the decomposition is now formed by five sets of values, 
\beqn
\sigma^{(1)}_{p_1}, \quad \sigma^{(2)}_{p_2},\quad \sigma^{(3)}_{p_3},\quad \sigma^{(4)}_{p_4},\quad\text{and } \sigma^t_{n},\label{c11}
\eeqn
which are also sorted in decreasing order.

Similarly to SVD,  without truncation
 the HOSVD  (\ref{c10}) is exact. Nevertheless, truncation is advised to filtering noise, spurious artifacts, or reducing the data-dimensionality depending on the need of our application. As in SVD, the number of modes retained for each case generally depends on a tolerance (tunable) as
\beqn
 \begin{split}
\sigma_{P_1+1}/\sigma_{1}\leq \varepsilon_{svd_1},\\
\sigma_{P_2+1}/\sigma_{1}\leq \varepsilon_{svd_2},\\
\sigma_{P_3+1}/\sigma_{1}\leq \varepsilon_{svd_3},\\
\sigma_{P_4+1}/\sigma_{1}\leq \varepsilon_{svd_4},\\
\sigma_{N+1}/\sigma_{1}\leq \varepsilon_{svd_5}.\\
\label{eq:TOLhosvd}
\end{split}
\eeqn
Generally, the tolerance is set the same for all the cases, so $\varepsilon_{svd_l}=\varepsilon_{svd}$, for $l=1,2,3,4,5$. Although, for highly complex dynamics, these tolerances should be set different, depending on the database studied. This is one of the main advantages of HOSVD algorithm compared to SVD, where the dimensionality of all the directions and components of the database is reduced similarly. Using HOSVD, it is possible to distinguish between different noise levels or components magniture.

After truncation, HOSVD (\ref{c10}) is written as 
 \beqn
 V_{j_1j_2j_3j_4k}\simeq\sum_{n=1}^{N} W_{j_1j_2j_3j_4n}\hat V_{kn},\label{c15}
\eeqn
where $W_{j_1j_2j_3j_4n}$ and $V_{kn}$ are the spatial and temporal modes, and $N$ is the spatial complexity defined above. The spatial modes are defined as 
   \beqn
 W_{j_1j_2j_3J_4n}= \sum_{p_1=1}^{P_1}\sum_{p_2=1}^{P_2}\sum_{p_3=1}^{P_3}S_{p_1p_2p_3p_4n}
 W^{(1)}_{j_1p_1} W^{(2)}_{j_2p_2} W^{(3)}_{j_3p_3} W^{(4)}_{j_4p_4}/\sigma^t_r,
 \quad \hat V_{kn}=\sigma^t_rT_{kn}.\label{c16}
  \eeqn
  


\subsection{Gappy SVD and gappy HOSVD for data repairing and resolution enhancement \label{sec:gappyRepairingGeneralSec}}

Repairing databases with corrupted or incomplete information, enlarging the dimension of the original database, and increasing its spatial resolution are some of the most relevant applications of SVD. In addition to the classical applications of SVD, patterns identification and reducing data dimensionality, when used properly, SVD is also a very useful tool for the post-processing and treatment of databases. The algorithm for this application is very simple and is based on the properties of the decomposition, which re-organizes the SVD modes as function of their contribution into the reconstruction of the original database. 

Starting from the sapshot matrix eq. (\ref{ab0}), with dimension $J\times K$, the algorithms for data repairing or resolution enhacement are presented below. For simplicity, the SVD-based algorithms introduced are particularized for a single snapshot, two-dimensional, organized in matrix form, while for three-dimensional databases and/or temporal variations, the algorithms uses HOSVD and the snapshot tensor (\ref{c43d}), or its corresponding version adapted to the number of components composing the tensor.

\subsubsection{Gappy SVD \label{sec:gappy}}
\begin{figure}[h!]
    \begin{center}
        \includegraphics[trim=0 100 0 100, clip, width=\textwidth]{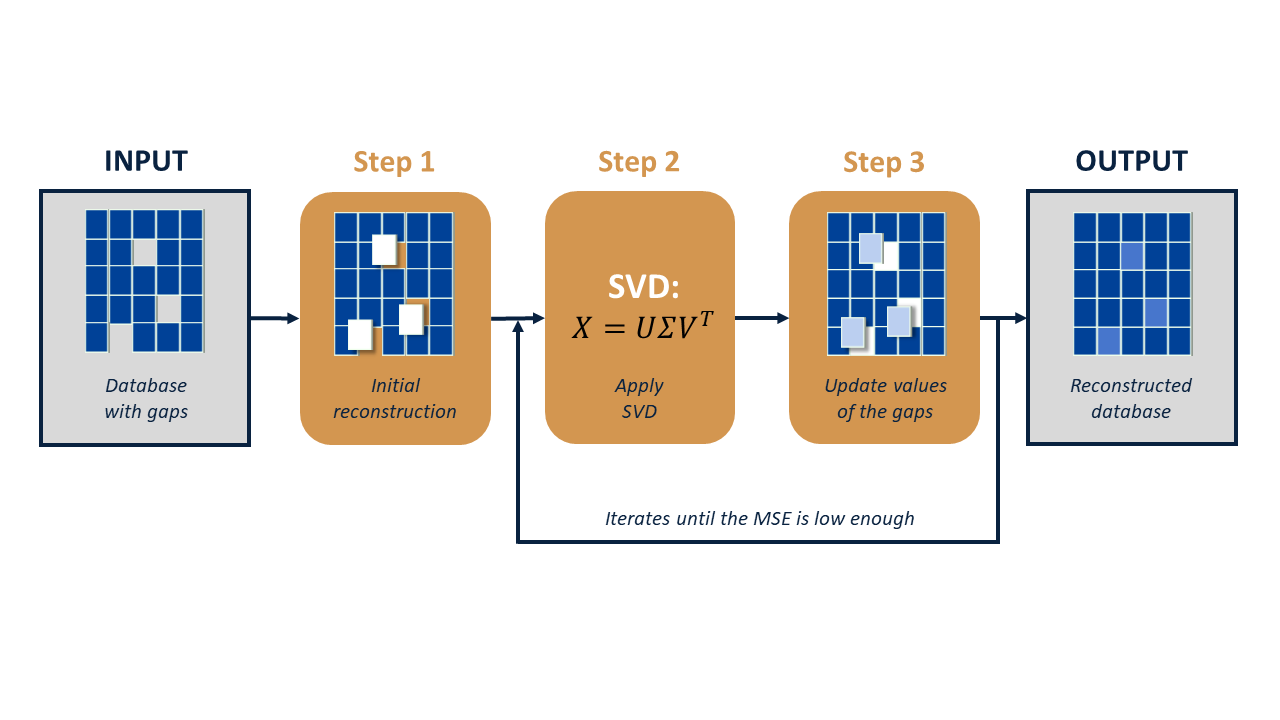}
        \vskip-0.1cm
    \end{center}
\vskip-0.5cm
\caption{Gappy SVD: sketch summarizing the methodology.\label{fig:gappy}}
\end{figure}

Gappy SVD, also known as Gappy POD, uses SVD iteratively to repair and reconstruct incomplete or corrupted databases. The database analysed is particularized for a single snapshot vector $\bv_k$ with dimension $J\times 1$, which is re-organized in matrix form, named as $\widehat\bV^{0}$ with dimension  $\widehat N_1\times\widehat N_2=J$, being $\widehat N_1$ and $\widehat N_2$ the dimensions associated to the streamwise and normal directions. The initial database contains some information that is corrupted, for instance, given by $NaN$ ({\it Not a Number}) information. To repair the database, Gappy SVD algorithm is as follows:
\begin{itemize}
    \item[\bf Step 1.] Initialize the database, $\widehat\bV^{0}$,  giving the points $NaN$ an initial value, which can be zero or can be calculated as the mean or as the linear or non-linear interpolation between the surrounded points. 
    \item[\bf Step 2.] Apply SVD to the previous matrix  $\widehat\bV^{i}=\bW^{i}\bSigma^{i}(\bT^{i})^T$, for $i=0$ in the initial iteration, and reduce the previous matrix dimensions by retaining $P'$ singular values (tunable).
    \item[\bf Step 3.] Reconstruct the new reduced snapshot matrix 
    \begin{equation}
       \widehat\bV^{i+1}=\bW_{P'}^i\bSigma_{P'}^i(\bT_{P'}^i)^T,     
    \end{equation}
    with $\bW_{P'}^i\in\mathbb{R}^{\widehat N_1\times P'}$,  $\bSigma_{P'}^i\in\mathbb{R}^{P'\times P'}$, and $\bT_{P'}^i\in\mathbb{R}^{\widehat N_2\times P'}$.
    \item[\bf Step 4.] Update the gaps with the values of $\widehat\bV_k^{i+1}$.
    \item[\bf Step 5.] Calculate the MSE at the gaps between iterations as
    \begin{equation}
        MSE_{gaps} = \dfrac{1}{N_{gaps}}\sqrt{\sum_{n=1}^{N_{gaps}} \lvert \widehat\bV^{i}-\widehat\bV^{i-1} \rvert},
    \end{equation}
    
    where $N_{gaps}$ is the number of gaps in the database. While $MSE_{gaps}<10^{-6}$, update $i=i+1$ and repeat steps 2--4 in the new matrix from Step 4.
\end{itemize}
The final reconstructed matrix $\widehat\bV_k^{i+1}$ (for iteration $i$) is the new repaired matrix. Figure \ref{fig:gappy} shows an sketch representing the general methodology. 

For three-dimensional data, the algorithm uses databases organized in tensor form, which particularized for a single snapshot has dimension $\widehat N_1\times\widehat N_2\times\widehat N_3$, with $\widehat N_3$ as the spanwise component. The previous algorithm is then the same, but SVD is replaced by HOSVD to properly repair all the components of the database. This algorithm is also valid for larger dimension databases, or it can be connected to other components of the database, different to the spatial dimension. See more details of the algorithm in Refs. \cite{Vega1HOSVDgappy,Venturietalgappy,Beckersetalgappy}.

\subsubsection{Increasing resolution using modal decomposition \label{sec:superResSVD}}
SVD can be used to increase the resolution of the database. The method uses SVD iteratively, and it also makes profit of the properties of SVD to re-organize SVD modes as function of their contribution into the original database. The algorithm is particularized for a single snapshot $\bv_k$, which is organized in matrix form $\bV^{DS,i}$ ($i$ is the iteration number of the algorithm) with (down-sampled) dimension $N_1\times N_2<J$, being $N_1$ and $N_2$ the dimensions associated to the streamwise and normal directions. The main goal of this algorithm is to get a new database better resolved in space, with dimension $\widehat N_1\times\widehat N_2=J$. The algorithm is as follows:
\begin{itemize}
    \item[\bf Step 1.] Apply SVD to the initial under-resolved or downsampled database,  and set the number of singular values to $P'$ (tunable), as
    \begin{equation}
        \bV^{DS,i}\simeq \bW_{P'}^{DS,i}\Sigma_{P'}^{DS,i}(\bT_{P'}^{DS,i})^T,\label{eq:svdDS}
    \end{equation}
    with $\bW_{P'}^{DS,i} \in \mathbb{R}^{N_1\times P'}$, $\bSigma_{P'}^{DS,i} \in \mathbb{R}^{P'\times P'}$ and $\bT_{P'}^{DS,i} \in \mathbb{R}^{N_2\times P'}$, for $i=0$ in the first iteration.
    \item[\bf Step 2.] Enlarge the dimension of the matrices from the previous decomposition using a linear (or non-linear) interpolation as:
    $\bW_{P'}^{DS,i+1} \in \mathbb{R}^{2N_1\times P'}$ and $\bT_{P'}^{DS,i+1}\in\mathbb{R}^{2N_2\times P'}$ interpolating between two points, and reconstruct the new enlarged matrix 
    \begin{equation}
        \bV^{DS,i+1}\simeq \bW_{P'}^{DS,i+1}\Sigma_{P'}^{DS,i}(\bT_{P'}^{DS,i+1})^T,\label{eq:svdDS2}
    \end{equation}
    \item[\bf Step 3.] Update the iteration number as $i=i+1$.
    \item[\bf Step 4.] Repeat steps 1 -- 3 $i=s$ times, until  $2^s \times N_1=\widehat N_1$ and $2^s \times N_2=\widehat N_2$. 
\end{itemize}

 The new matrix $\bV^{DS,s+1}$ is the matrix with enhanced resolution.

 When working with three-dimensional data, the algorithm employs tensor databases that are organized in a specific way. For a single snapshot, the database dimension is $\widehat N_1\times\widehat N_2\times\widehat N_3$, where $\widehat N_3$ represents the spanwise component. To refine this process further, the previous algorithm is modified by replacing SVD with HOSVD. Also, when dealing with temporal components, HOSVD is used to enhance the resolution of the databases, and the associated temporal information (tunable) can either remain constant withing the algorithm, only enhancing the resolution of the spatial components, or may also be enlarged, interpolating to time instants that were not included in the original dataset.
 
See more details regarding this algorithm and some applications in Refs. \cite{VegaSVDResolution,NoureletalNMR}

\subsection{Higher order dynamic mode decomposition (HODMD)\label{sec:HODMD}}

Higher order dynamic mode decomposition (HODMD) \citep{LeClaincheVega17} is an extension of dynamic mode decomposition (DMD) \citep{Schmid10} introduced for the analysis of complex flows and highly non-linear dynamical systems \citep{VegaLeClaincheBook20}. 

HODMD decomposes spatio-temporal data $\bv_k$,  as an expansion of $M$ DMD modes $\bu_m$, weighted by their corresponding amplitude $a_m$ as
    \begin{equation}
 \bv(x,y,z,t_{k})\simeq  
 \sum_{m=1}^M a_{m}\bu_m(x,y,z)e^{(\delta_m+i \omega_m)t_k},\label{ab00}
  \end{equation}
for $k=1,\ldots,K$, where $\omega_m$ is the oscillation frequency and $\delta_m$ corresponds to the growth rate, representing the mode temporal growing, decaying or showing when the mode remains neutral in time. 

HODMD algorithm is briefly introduced here, where the method is summarized in two main steps. A detailed description of the algorithm can be found in Ref. \cite{LeClaincheVega17}. We recommend the book by Vega \& Le Clainche \cite{VegaLeClaincheBook20}, where it is possible to find a wide range of examples and applications, as well as the implementation of the algorithms in Matlab \cite{matlab}.

The algorithm is presented as follows.
\begin{itemize}
\item {\bf Step 1: Dimension reduction via SVD.} 
SVD is applied to the snapshot matrix (\ref{ab0}) to remove noise or spatial redundancies and reduce data dimensionality from the spatial dimension $J$ to the spatial complexity $N$ (number of SVD modes retained). At this step, tolerance $\varepsilon_{svd}$ is fixed to select $N$.

Starting from eq. (\ref{ab20}), it is possible to define the {\it reduced  snapshot matrix} as
\begin{equation}
\widehat{\bV}_1^K=\bSigma\,\bT^T, \label{ab22}
 \end{equation}
with $\bV_1^K=\bW\widehat{\bV}_1^K$ and $\widehat{\bV}_1^K$ with dimension $N\times K$.

\item{\bf Step 2: The DMD-d algorithm.}
The {\it high order Koopman assumption} is applied to the reduced snapshot matrix as
\beqn
\widehat{\bV}_{d+1}^K\simeq \widehat{\bR}_1 \widehat{\bV}_1^{K-d}+ \widehat{\bR}_2 \widehat{\bV}_2^{K-d+1} + \ldots + \widehat{\bR}_d \widehat{\bV}_d^{K-1},\label{ab26}
\eeqn
where $\widehat{\bR}_k=\bW^T\bR_k\bW$ for $k=1,\ldots,d$ are the Koopman operators, linear, which contain the system dynamics. The snapshot matrix eq. (\ref{ab0}) is then divided into $d$ blocks formed by $K-d$ time-delayed snapshots each. For $d=1$, the method is similar as standard DMD \cite{Schmid10}.

The previous equation is reorganized in the following way
\beqn
\left[\begin{array}{c}\widehat{\bV}_2^{K-d+1}\\
\ldots\\ \widehat{\bV}_{d}^{K-1}\\\widehat{\bV}_{d+1}^{K} \end{array}\right] =
\left[\begin{array}{cccccc}
\bo&\bI &\bo&\ldots &\bo&\bo\\
\bo &\bo&\bI&\ldots&\bo &\bo\\
\ldots & \ldots & \ldots&\ldots&\ldots&\ldots \\
 \bo &\bo&\bo& \ldots&\bI &\bo\\
 \widehat{\bR}_1 &\widehat{\bR}_2&\widehat{\bR}_3& \ldots&\widehat{\bR}_{d-1}&\widehat{\bR}_d \end{array}\right]
\cdot 
\left[\begin{array}{c}\widehat{\bV}_1^{K-d}\\ \widehat{\bV}_2^{K-d+1}\\
\ldots\\ \widehat{\bV}_{d}^{K-1} \end{array}\right].
 \label{ab32}
\eeqn
The previous expression can also be represented as
\beqn
\tilde{\bV}_2^{K-d+1}=\tilde{\bR}\,\tilde{\bV}_1^{K-d},\label{ab30}
\eeqn
where $\tilde{\bV_1}^{K-d+1}$ and $\tilde{\bR}$ correspond to the {\it modified snapshot matrix} and {\it modified Koopman matrix}, respectively. SVD is again applied to this matrix to remove expected spatial redundancies, where the tolerance $\varepsilon_{svd}$ is again used to retain $N'>N$ SVD modes, as
\beqn
\tilde\sigma_{N'+1}/\tilde\sigma_{1}<  \varepsilon_{svd}.
\label{ab37}
\eeqn
It is remarkable that the key of the good performance of HODMD algorithm lies in matrix $\tilde{\bR}$, which increases the spatial complexity of the data from $N$ to $N'$.

The next step calculates the DMD modes to form the DMD expansion eq. (\ref{ab00}). These modes are obtained via solving an eigenvalue problem in the new reduced version of the modified Koopman matrix, whose eigenvalues and eigenvectors are adapted to represent the DMD frequencies and growth rates, and the DMD modes, respectively. Finally, the mode amplitudes are calculated via least-squares fitting, following the optimized DMD method \cite{Chenetal2012}. It is remarkable that depending on the way the amplitudes are calculated, the mode weight can be different. See for instance HODMD with criterion (HODMDc) \cite{KouLeClaincheZhangPoF18}, where the modes also consider the contribution of the growth rates in the amplitude calculations. A detailed description on different ways to calculate the DMD mode amplitudes can be found in the review paper Ref. \citep{BekaetalPoF2023}.

Finally, sorting the mode amplitudes in decreasing order, the DMD modes are reordered. The $M\leq N'$ retained DMD modes are calculated based on a second tolerance $\varepsilon_{a}$ (tunable) as
 \beqn
 a_{M+1}/a_1<\varepsilon_{a}. \label{b66}
 \eeqn
$M=min(K,N')$ is known as the {\it spectral complexity}, different from the {\it spectral dimension} $K$. 
For a sufficiently large number of snapshots $K$ (which is the common case), in cases in which the spatial complexity $N$ is smaller than the spectral complexity $M$ (for $K> N$), the high order Koopman assumption completes the lack of spatial information (reduced to $N$) and ensures the good performance of the DMD method. When using standard DMD, it is possible to find some cases in which $N<M$, hence the algorithm fails. HODMD in contrast, overcome such limitation, extending the range of application of the standard algorithm. HODMD method has shown potential in the study of various flow types such as transition to turbulent \citep{LeClaincheVegaSoria17,LeClaincheEnergies19} and turbulent flows \cite{LeClaincheetalJFM20,LeClaincheetalJFM22}, identification of flow features from experimental data with noise \citep{LeClaincheVegaSoria17,LeClaincheetalAIAA17,LeClaincheetalJAircraft18}, in the analysis of data obtained from limited spatial locations like field measurements \citep{LeClaincheetalJAircraft18,LeClaincheetalEnergies18}, or even in medical imaging \cite{NoureletalHODMD}.
\end{itemize}

\subsection{Multi-dimensional iterative higher order dynamic mode decomposition {\em(mdHODMD-it)}}

The multi-dimensional HODMD (mdHODMD) algorithm  uses the snapshot tensor given in equation (\ref{c43d}). So, instead of SVD, HOSVD is applied as first step to reduce data-dimensionality. This method considers the variables and components of the database separately, so different modes can be selected to better represent each fiber forming the tensor.

The  algorithm is summarized in two steps:
\begin{itemize}
\item {\bf Step 1:} Application of HOSVD.
	\begin{itemize}
\item[1.1] Perform HOSVD on the snapshot tensor $\bV$, eq. (\ref{c43d}), without truncation. The ranks of the matrices, whose columns are the fibers, determine the number of modes in the tensor $\bV$. These ranks are $P_1=\min{J_1,J_2J_3J_4K}$, $P_2=\min{J_2,J_1J_3J_4K}$, $P_3=\min{J_3,J_1J_2J_4K}$, and $N=\min{K,J_1J_2J_3J_4}$. The resulting singular values are selected as shown in equation (\ref{c11}). 
\item[1.2] Choose spatial and temporal tolerances, denoted as $\varepsilon_{svd}$ and $\varepsilon_a$ respectively, to determine the number of modes retained in each direction. These numbers are denoted by $P_1$, $P_2$, $P_3$, $P_4$ and $N$. The smallest values of $P_1$, $P_2$, $P_3$, $P_4$ and $N$ that satisfy the condition given by equation (\ref{eq:TOLhosvd}) are chosen. 
\item[1.3] Perform truncated HOSVD with the numbers of modes determined in the previous step. This gives the core tensor and the modes that define the truncated HOSVD on the right-hand side of equation (\ref{c10}). 
\item[1.4] Compute the spatial and temporal modes as defined in equation (\ref{c16}). 
\end{itemize}
\item {\bf Step 2:} Calculation of the multidimensional DMD expansion.
Step 2 in the HODMD algorithm described in
 \S\ref{sec:HODMD} is  applied to the reduced snapshot matrix defined by the temporal modes $\hat V_{kr}$ calculated
in the previous step item 1.4. The mode amplitudes, frequencies, growth rates and DMD modes are calculated at this step. 
\end{itemize}

When the data are too noisy or too complex, mdHODMD can be applied iteratively, and the algorithm is called as multi-dimensional iterative HODMD (mdHODMD-it). More specifically, as firs step the mdHODMD technique is reapplied to the reconstructed snapshots with the same tolerances as in the first application, resulting in cleaner data. 
Next, the algorithm is applied to the newly reconstructed snapshots, and this process repeats until the number of HOSVD modes is maintained after two consecutive iterations.  With each iteration, the algorithm recalculates and orders both HOSVD and DMD modes based on their new corresponding amplitudes, ultimately improving the quality of the DMD reconstructions. The primary benefit of the iterative approach is the removal of irrelevant or inconsistent HOSVD modes based on the tolerance $\varepsilon_{svd}$.

\section{Module 2: hybrid machine learning tools\label{sec:module2}}

Module 2 is formed by a group of deep learning algorithms that are combined with the modal decomposition methods presented in Module 1. Modal decomposition are suitable to identify patterns that contain relevant information about the physics of the dynamical system. As previously mentioned, these techniques have excellent properties to reduce the data-dimensionality, which is very convenient in fluid dynamics, other complex problems and industrial applications. The dimensionality of the original database is reduced from hundred thousands or even millions degrees of freedom to a few modes, POD or DMD, generally varying from dozens to hundred, depending on the problem understudy. These modes represent the dynamics of the system, which is represented by modal decomposition with physical interpretability. As second step, the reduced database is combined with different algorithms of deep learning. ModelFLOWs-app in particular offers two possibilities, convolutional neural networks (CNNs) and recurrent neural networks (RNNs), although the algorithm is opened to be combined with other types of architectures. This so-called hybrid ROM, can be used for several applications, including data repairing and reconstruction (Module 2 - application 2) and temporal forecasting (Module 2 - application 3), which is used to predict the non-linear dynamics in complex systems. This application is able to predict saturated (converged or statistically steady) solutions from transient stages in numerical simulations, which generally results in a notorious reduction of computational times to generate new databases. In the examples presented, the computational cost is reduced from hundred or even thousands of computational hours to a few seconds \cite{AbadiaetalExpSystAppl22}.

Module 2 - application 1 also explores the possibility of using deep learning architectures to identify patterns in the dynamical system understudy. For such aim, several architectures of  autoencoders have been tested, and compared with results obtained with modal decompositions. This application completes the Deep Learning Module, which is complementary to the modal decomposition module. Module 1 and Module 2 solves similar applications using classical (modal decomposition) and more modern (neural networks) tools, respectively.

\subsection{Dimensionality reduction and via autoencoders (AEs)}\label{subsec:module2_AEs}

\begin{figure}[h]
    \centering
    \includegraphics[trim=100 50 130 50, clip, width=\textwidth]{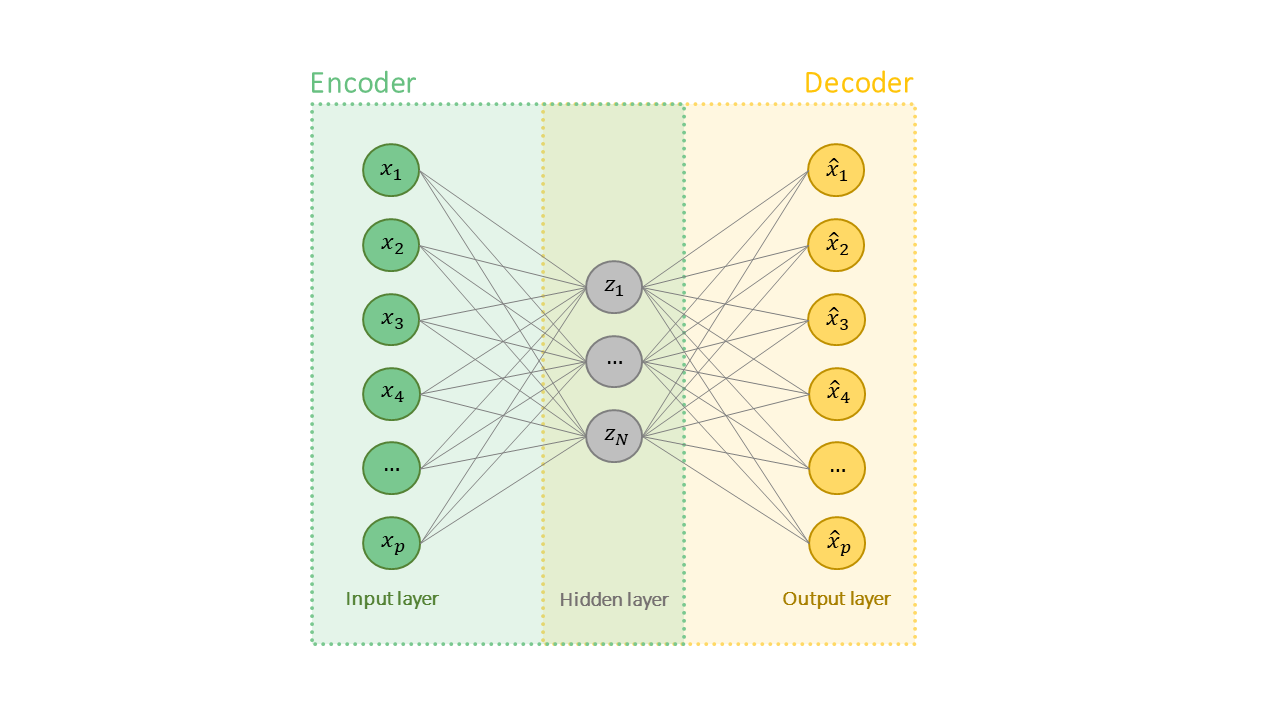}
    \caption{Basic architecture of autoencoders. \label{fig:AE}}
\end{figure}

Autoencoders (AEs) are unsupervised neural networks designed to learn a compressed representation of a database. The encoder is responsible for projecting the data onto a low-dimensional nonlinear manifold,
 $\bv \mapsto \boldsymbol{r}$,
while the decoder reconstructs the data from the latent space back to the reference space and reduces the reconstruction error, See Fig. \ref{fig:AE}

By training the autoencoder model, it acquires the ability to identify the crucial features in the data necessary for reconstruction. This is accomplished by optimizing the model parameters $\boldsymbol{w}$ to minimize the reconstruction loss $\mathcal{L}_\mathrm{rec}$. For $\mathcal{E}$ and $\mathcal{D}$ being the encoder and decoder respectively, this can be expressed as,
\begin{subequations}
    \begin{equation}
        \mathcal{F} = \mathcal{D} \circ \mathcal{E}, 
    \end{equation}
    \begin{equation}
        \widehat\bv = \mathcal{F}(\bv; \boldsymbol{w}), \\
    \end{equation}
    \begin{equation}
        \mathcal{L}_\mathrm{rec} = \epsilon(\bv, \tilde\bv), 
    \end{equation}
\end{subequations}
being  $\epsilon$, $\bv$ and $\tilde\bv$  the loss function, the input data, and the reconstruction of the input data, respectively.

The simplest form of AEs is formed by non-recurrent, feed-forward neural networks that comprise an input layer, one or more hidden layers, and an output layer with the same number of neurons as the input layer. Their primary objective is to reconstruct the input data, minimizing the difference between input and output through a loss function that can accommodate regularization and sparsity terms. ModelFLOWs-app uses this type of AEs to identify patterns in the data analysed. Different types of activation functions, number of layers, or compression rates, among other parameters, can be selected. Snapshot matrix eq. (\ref{ab0}) is the input of the AE architecture used, and the output provides the reconstructed matrix using a specific (tunable) number of AEs, as well as the patterns collected in each AE. These AE are re-organized in decreasing order as function of their highest influence in the reconstructed field, minimising the RRMSE reconstruction error eq. (\ref{eq:rrmse}).

It is important to note that a shallow AE with linear activation functions is equivalent to principal component analysis (PCA), an extension of POD. Hence, in this case, similarities are found between AE modes and POD (or SVD) modes. Nevertheless, in the context of modal decomposition, the autoencoder architecture offers an appealing framework that can effectively incorporate non-linearity in mappings by utilizing non-linear activation functions \cite{ae_modal}.


\subsection{Hybrid reduced order models\label{sec:hybridROM}}

\begin{figure}[H]
    \centering
    \includegraphics[trim=0 50 0 50, clip, width=0.95\textwidth]{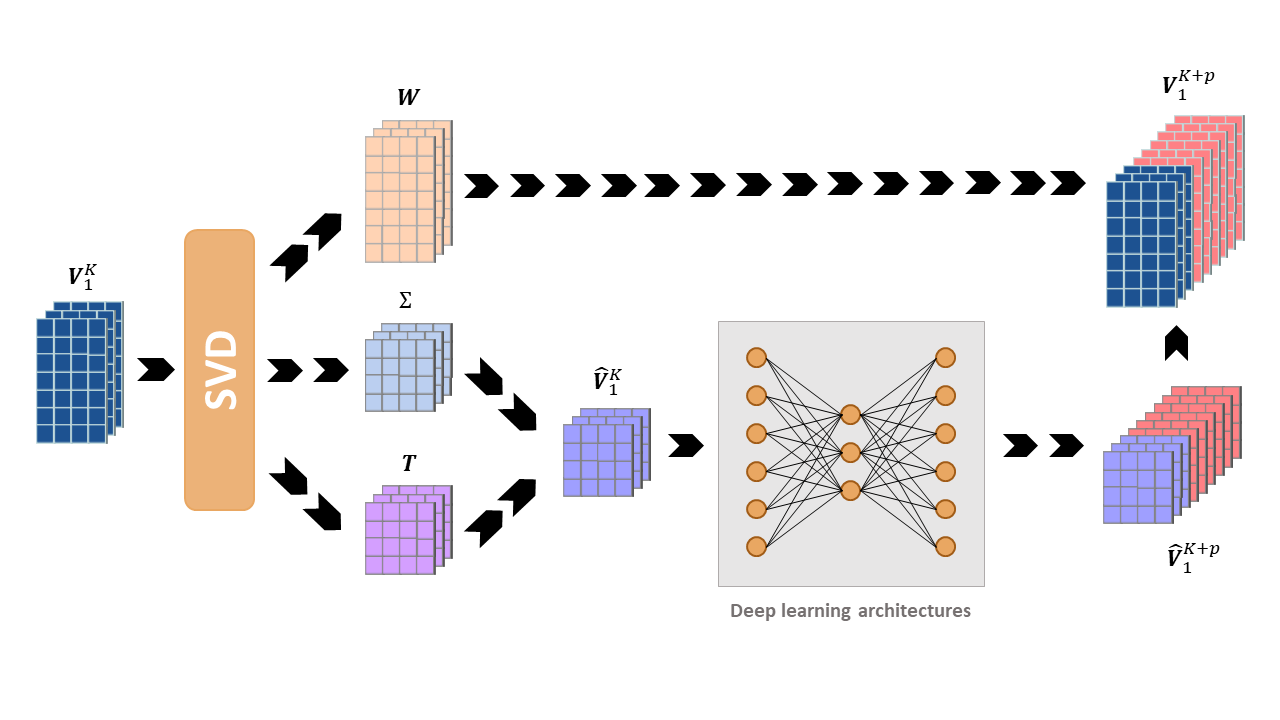}  
    \caption{Steps involved in the architecture of the hybrid reduced order models.}
    \label{fig:HybridROM}
\end{figure}

Similarly to HODMD algorithm section \ref{sec:HODMD}, hybrid ROMs are defined by two main steps. The first step starts from eq. (\ref{ab20}), for simplicity repeated here as
\begin{equation}
\bV_1^{K}\simeq\bW\,\bSigma\,\bT^\top,\label{ab20v2}
\end{equation}
to define the {\it reduced  snapshot matrix} (with dimension $N\times K$) as
\begin{equation}
\widehat{\bV}_1^K=\bSigma\,\bT^T, \label{ab22}
 \end{equation}
 
with $\bV_1^K=\bW\widehat{\bV}_1^K$.
The second steps applies deep learning architectures into this reduced matrix. Depending on the type of architecture, it is possible to predict temporal evolution of a signal  or to reconstruct two- or even three-dimensional databases from sensors, as presented in sections  \ref{sec:predictionDNN} and \ref{sec:reconstructionDNN}, respectively. Figure \ref{fig:HybridROM} shows an sketch summarizing the hybrid ROM methodology for temporal foreccasting, where $p$ snapshots are predicted. So, initial snapshot matrix $\bV_1^K$ eq. (\ref{ab0}) is then transformed to the new snapshot matrix containing $p$ additional snapshots $\bV_1^{K+p}$.

It is remarkable that using this method, one-dimensional architectures are required, since they are directly applied to the reduced snapshot matrix. In contrast, two- or even three-dimensional architectures should be applied to the original snapshot matrix, which also contains relevant spatial information of the database. Hence, the reduction in computational time and memory of this hybrid methodology is one of the key points of the algorithms presented.

The performance of the architectures presented below is optimal for standard fluid dynamics problems. However, when the complexity of the database increases, it is necessary to take into account the magnitude and variance of the variables analysed. For instance, in reactive flows, where a large number of variables are involved represeting the multiple species of the flow (more than $80$), it is necessary to combine the previous methodology with other pre-processing techniques, such as centering and scaling. These two methods are applied to each one of the multiple variables of the database before forming the snapshot matrix $\bV_1^K$ eq. (\ref{ab0}). Centering removes the temporal mean in each variable, in this way the analysis only considers the fluctuation field. Scaling normalizes all the variables so they can be compared on the same bases. There are several possibilities to scale the data. The following equation represents centering and scaling of the data, 
\begin{equation}
	\tilde{\bv}_j(t_k) = \dfrac{\bv_j(t_k)-\bar{v}_j}{c_j}\label{e24},
\end{equation}
where $\bv_j$ is the \textit{j-th} variable, $\bar{v}_j$ is the mean averaged in time, $c_j$ is the scaling factor used and $\tilde{\bv}_j$ is the scaled variable. Three scaling techniques have been implemented in ModelFLOWs-app:

\begin{enumerate}
	\item \textit{Range} scaling: The difference between the maximum and minimum value of each variable is used as the scaling factor.
	\item \textit{Auto} scaling: the standard deviation of the $j$-th variable ($\sigma_{j}$) is used as the scaling factor.
	\item \textit{Pareto} scaling: the square root of the standard deviation of the $j$-th variable ($\sqrt{\sigma_{j}}$) is used as the scaling factor.
\end{enumerate}

Several studies in the literature \cite{PCA,d2020analysis,Corrochano22} have  examined the impact of centering and scaling on modal decomposition techniques like PCA and HODMD in the context of combustion applications.

\subsubsection{Time-series forecasting models using convolutional and recurrent neural networks (CNN and RNN)\label{sec:predictionDNN}}

The proposed architectures have been chosen to best approximate temporal evolution of databases with high-dimensional nature of time series driven by strongly sequential dynamics. The architectures are fit to predict the longest data sequence as possible, intending to minimize the quantity of training data. In this way, the predictive hybrid ROM can be used to predict new databases in numerical simulations, generally associated to large computational costs. Also, this strategy avoids overfitting.

\begin{figure}[H]
	\centering
	\includegraphics[trim=80 70 80 50, clip,height=0.6\columnwidth]{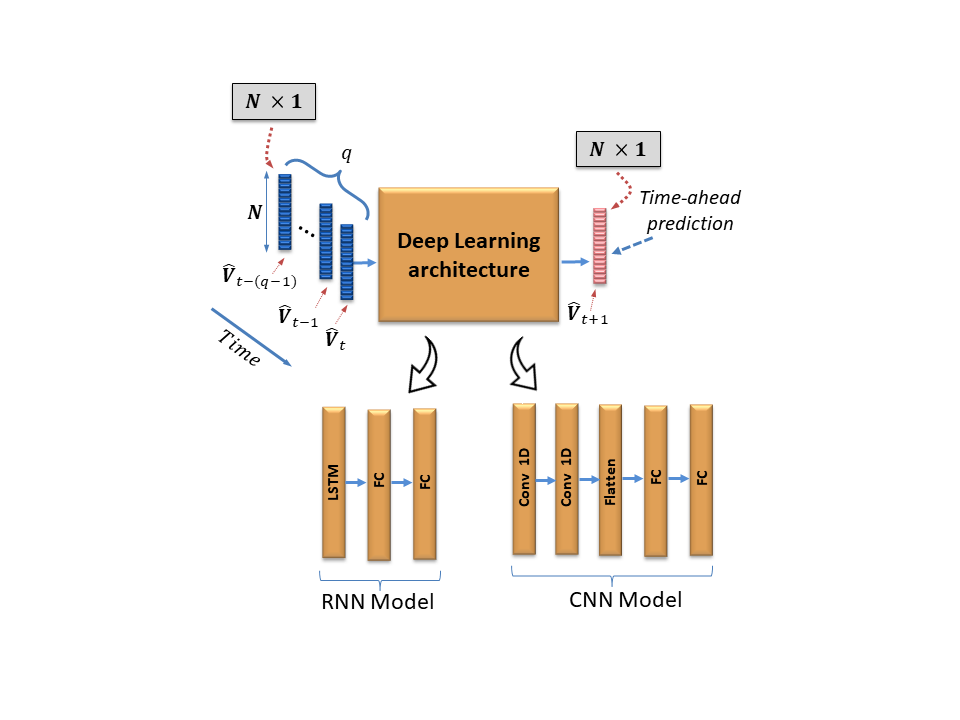}
    \caption{Hybrid predictive ROM. Recurrent and convolutional neural networks to predict a time-ahead snapshot, $\widehat{\bV}_{k+1}$, based on the previous $q$ snapshots. N is the number of POD modes. 
    \label{fig:ML1}}
\end{figure}

 Fig. \ref{fig:ML1} shows an sketch showing this predictive deep learning models.  Each column of matrix $\widehat{\bV}_{k+1}$ represents the temporal dynamics of each snapshot. 
The model predicts the snapshot $k+1$, given at time $t_{k+1}$ in the reduced matrix, $\widehat{\bV}_{k+1}$, by utilizing data from the $q$ previous snapshots, defined as $\widehat{\bV}_{k}$, $\widehat{\bV}_{k-1}$, $\cdots$, $\widehat{\bV}_{k-q+1}$.

The model using CNN architectures is composed by one-dimensional convolutional (Conv 1D) layers followed by FC layers. Conv 1D applies a one dimensional kernel, with no padding and a stride of 1 \citep{Kiranyazaetal2021}. Flatten function is included in thee intermediate layers that bridge the gap between convolutional and FC layers, to transform the matrix structure employed by Conv 1D into a vector structure compatible with the FC layers.

The RNN model is formed by short-term memory (LSTM) layers and FC layers \citep{Yuetal19}. Different dimensions of the output space (number of units) for the LSTM layers have been considered to form this model. Although, in ModelFLOWs-app, different architecture details can be considered for this application as function of the problem analysed. Details about the number and type of layers, activation function, layer dimensions, ..., are presented in some examples in section \ref{sec:ResPredictionNN}. Hyper-parameter approach to automatically set these parameters to get the best predictions have also been implemented in ModelFLOWs-app. This function also set the optimization \citep{Kingmaetal} and learning rate parameters, batch size and number of epoch, used with early stopping method.

Before entering the deep learning models, the reduced snapshot matrix $\widehat{\bV}_1^K$ can be scaled to improve the performance of the neural network. Some of them have been already mentioned in the previous section. Apart from them, a new scaling method (\textit{Max per Mode (MpM)}) has been implemented, in which each column of $\widehat{\bV}_1^K$ ($\bv_j$) is scaled with the sum of the maximum values of all columns, as
\begin{equation}
	\hat{\bv_j} = \dfrac{\bv_j}{\sum_{j = 1}^{N}\text{max}|\bv_j|}\label{eq:Kapta}.
\end{equation}
This scaling method has been proved suitable for the prediction of non-periodic temporal modes \cite{Kaptanoglu21,paperAdricombustion}.

\begin{figure}[H]
	\centering
	\includegraphics[trim={110 95 95 110}, clip,height=0.5\columnwidth]{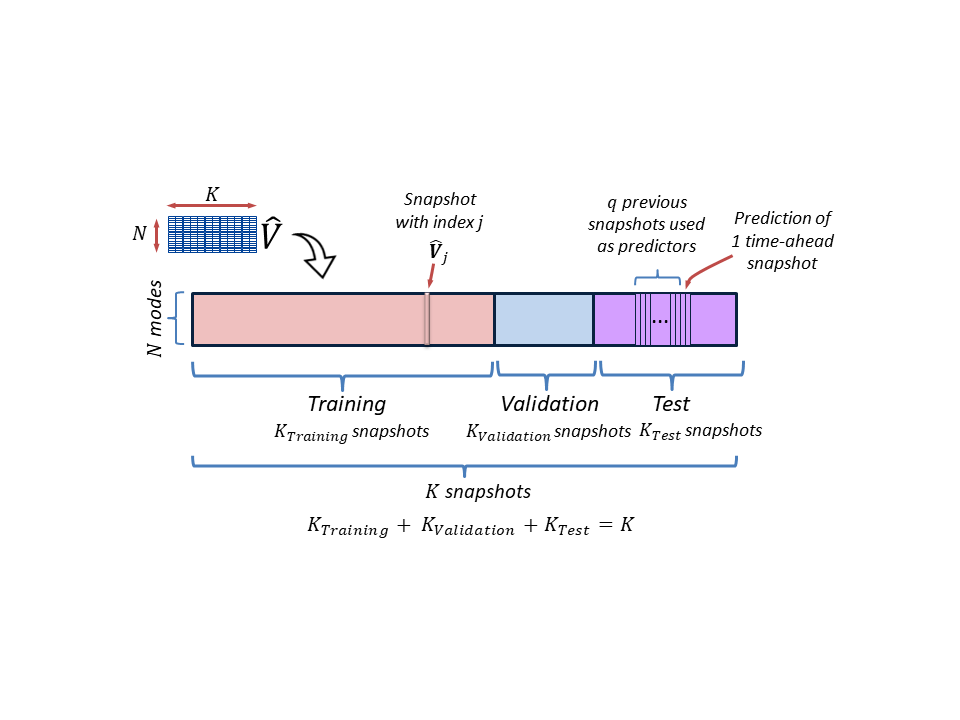}  
    \caption{Sketch with the training, validation and test set distribution for the deep learning module. \label{fig:ML2}}
\end{figure}

For the training and validation,  the columns $K$ of matrix $\widehat{\bV}$ are separated in three blocks with dimension $N \times K_{training}$, $N\times K_{validation}$ and $N\times K_{test}$, for the training, validation and test set, respectively, where $K=K_{training}+K_{validation}+K_{test}$. An sketch is presented in Fig. \ref{fig:ML2}. 

\begin{figure}[H]
	\centering
	\includegraphics[trim=100 80 100 80, clip,width=0.5\textwidth]{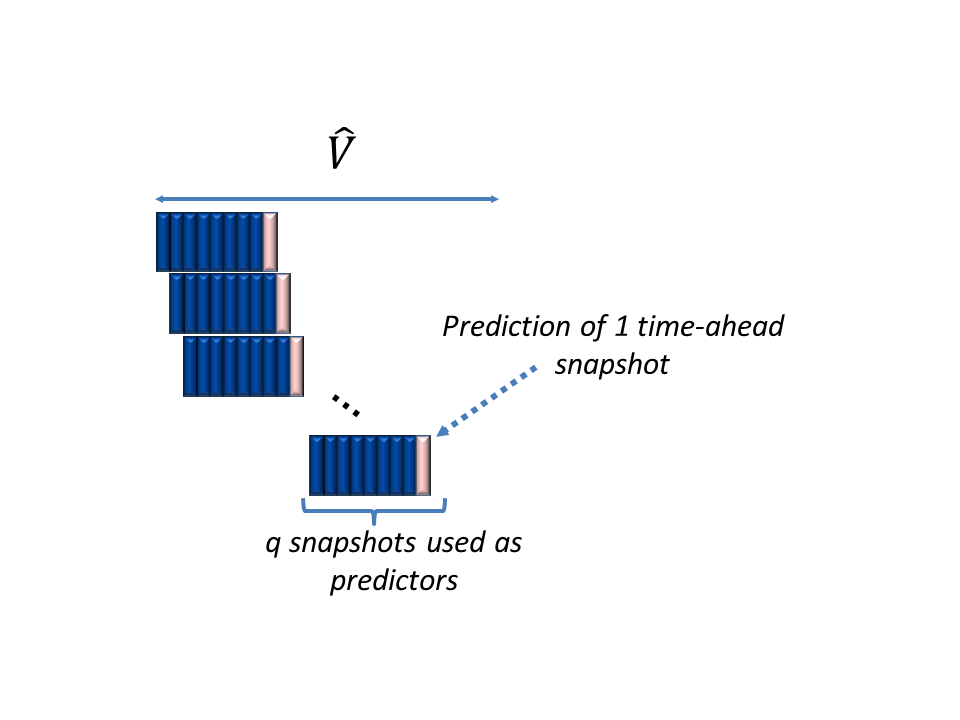}  
    \caption{Rolling window method calculating  $1$ output from $q$ inputs. \label{fig:ML3}}
\end{figure}

Similar to HODMD algorithm (section \ref{sec:HODMD}), a rolling-window method is used. The method generates one output, the model prediction, for $q$ inputs  (model information input) organized in data batches. As seen in the sketch of Fig. \ref{fig:ML3},  $1$ is the offset considered between the successive rolling windows.

Mean Squared Error Loss ($MSE_{Loss}$)  comparing the real and predicted databases ($\widehat{\bV}_t^{real}$, $\widehat{\bV}_t^{predicted}$) is minimized using batch stochastic gradient descent algorithm. The global loss ($MSE_{Loss}$) is based on averaging in time the local loss, which is calculated for each prediction as
\begin{equation}
MSE_{Loss}(t)=\frac{1}{N}||\widehat{\bV}_t^{predicted}-\widehat{\bV}_t^{real}||^2,\label{eq:MSEloss}
\end{equation}
being $N$  the number of singular values. The local error is calculated at the end of each epoch,  and early stopping method is used in the validation set to get the best network parameters. See more details of this architecture in Ref. \cite{AbadiaetalExpSystAppl22}.

Finally, the RRMSE eq. (\ref{eq:rrmse}) is also computed comparing the predictions with the real solution in the neural networks.

\subsubsection{Increasing resolution using hybrid deep learning models\label{sec:reconstructionDNN}}

\begin{figure}[H]
	\centering
        \includegraphics[width = 0.95\textwidth]{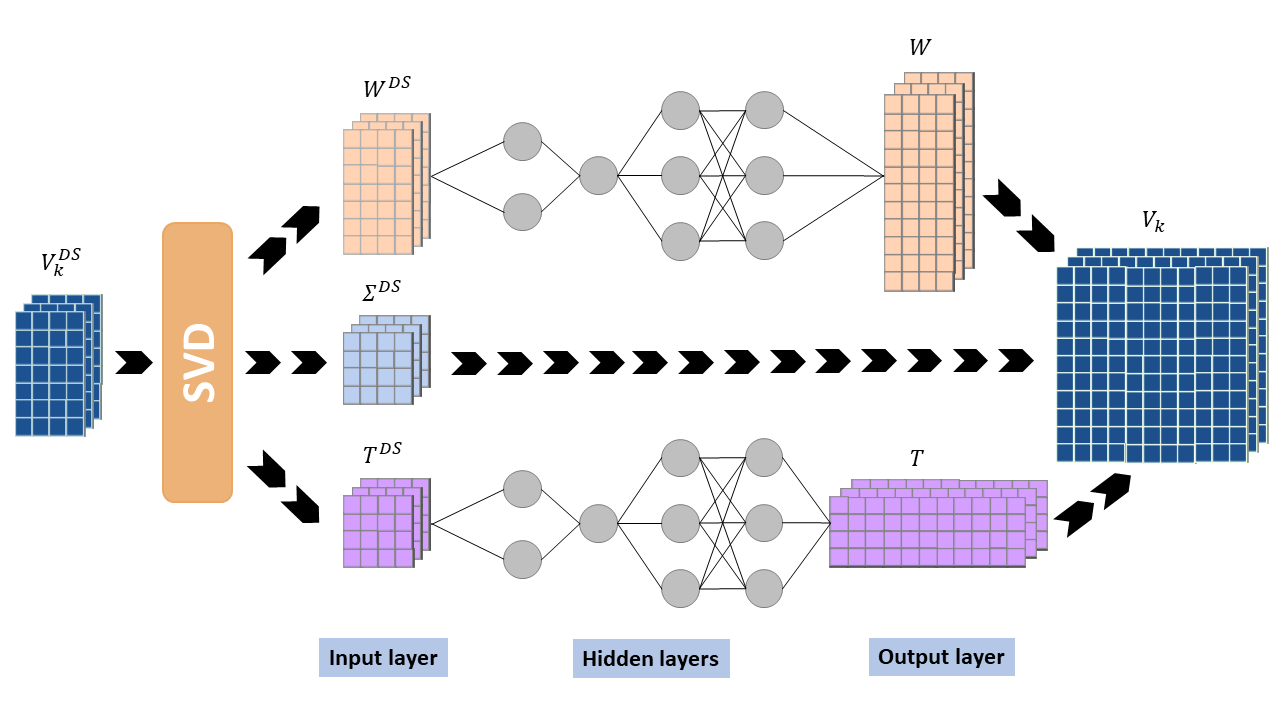}
	\caption{Reconstruction of databases combining SVD and deep learning. Sketch of the methodology. \label{fig:NN_arq}}
\end{figure}

This application reconstructs two- or three-dimensional databases using the information contained in a few sensors. As previously mentioned, the first step of the algorithm applies SVD to the snapshot matrix as defined in eq. (\ref{ab20v2}). In this case, the snapshot matrix $\bV_1^K$ eq. (\ref{ab0}), is now called as $\bV^{DS}$, as it represents a database under-resolved, only formed by a few points where temporal information about velocity vector, pressure field or other quantities are collected in each one of such points. For for each temporal snapshot, the dimensions of the {\em down-sampled} snapshot matrix $\bV^{DS}$ are  $N_1\times N_2$, being $N_1 \times N_2 <J$. For two-dimensional databases, $N_1$ and $N_2$ correspond to the streamwise and normal components, respectively, while for three-dimensional databases, $N_2$ collects the information of the normal and spanwise components (although the three spatial components could be re-organized differently into $N_1$ and $N_2$ based on our needs of filling gaps into the databases). Eq. (\ref{ab20v2})  particularized for a single snapshot, re-organizes the information into the down-sample matrix and can be re-written as 
\begin{equation}
\bV^{DS}\simeq\bW^{DS}\,\bSigma^{DS}\,(\bT^{DS})^\top.\label{ab20v3}
\end{equation}
The number of singular values retained is $P'$, hence $\bSigma^{DS}\in \mathbb{R}^{[P',P']}$.
Matrices $\bW^{DS}\in \mathbb{R}^{[N_1,P']}$ and $\bT^{DS}\in \mathbb{R}^{[N_2,P']}$, are introduced into a deep learning architecture to enlarge the dimension of $N_1$ and $N_2$ as in the original database, to $\widehat N_1$ and $\widehat N_2$, with $\widehat N_1 \times \widehat N_2 =J$, being the dimension of these new enlarged matrices as $\bW\in \mathbb{R}^{\widehat N_1,P'}$ and  $\bT\in \mathbb{R}^{\widehat N_2,P'}$. Once these new matrices have been modelled, they are again combined to re-construct the database with spatial dimension $J$ for each snapshot $K$, as
\begin{equation}
\bV_k\simeq \bW\,\bSigma^{DS}\,(\bT)^\top,\label{ab20v4}
\end{equation}
where $\bV_k \in \mathbb{R}^{[\widehat N_1,\widehat N_2]}$. 
The deep learning architectures gives this reconstructed matrix as an output. This architecture is formed by two groups of neural networks, defined as decoders (the decoding part of an autoencoder)  working in parallel that meets in the output layer, where the reconstructed solution is calculated as in eq. (\ref{ab20v4}). Hence, the  reconstruction error (RRMSE, see eq. (\ref{eq:rrmse})) used to improve the weights is calculated by comparing the reconstructed database with the original solution. Fig. \ref{fig:NN_arq} shows an sketch of the present architecture, formed by $5$ layers, although this is a tunable parameter of ModelFLOWs-app.
To organize the database, the strategy followed divides the total number of available snapshots into training, validation and test, as in Fig. \ref{fig:ML2}. It is also notorious that this architecture is able to predict in time the reconstruction of databases. 

See more details about this architecture and the model previsouly described in Ref. \cite{paperPaula}.

\section{Results\label{sec:results}}

\subsection{Module 1 - application 1: patterns identification}

HODMD algorithm (Sec.~\ref{sec:HODMD}) has been applied to identify temporal patterns in complex flows.  The good performance of HODMD and mdHODMD-it algorithms depends on the selection of the parameter $d$, which defines the number and sizes of the window selecting sub-areas of the analyzed matrix, and the tolerances $\varepsilon_{svd}$ and $\varepsilon_a$. As explained in Ref. \cite{LeClaincheVega17}, as a reference $d$ should be set in the interval $K/10<d<K/2$. There is not only a single optimal value for $d$. Generally, robust results are obtained for several values. Also, in saturated flows or converged signals, robust results should be obtained using similar values of $d$ applying the algorithm in different time intervals. The optimal values of $d$ also varies with the tolerances $\varepsilon_{svd}$ and $\varepsilon_a$, the time interval $\Delta t$ and the number of snapshots $K$. Also $d$ scales with the total number of snapshots, so if the $K$ is doubled, $d$ should also be multiplied by $2$. Finally, the tolerances $\varepsilon_{svd}$ and $\varepsilon_a$ should be comparable to the expected uncertainty of the measurements. If the level of noise of the database is known, $\varepsilon_{svd}$ should be set similar or slightly smaller (when the iterative method is applied) to such level. More details on the calibration of the algorithm can be found in Refs. \cite{LeClaincheVegaSoria17,VegaLeClaincheBook20,LeClaincheetalJFM20}.

\begin{figure}
    \centering
         a)
	\includegraphics[trim=150 100 220 100, clip, width=0.26\textwidth]{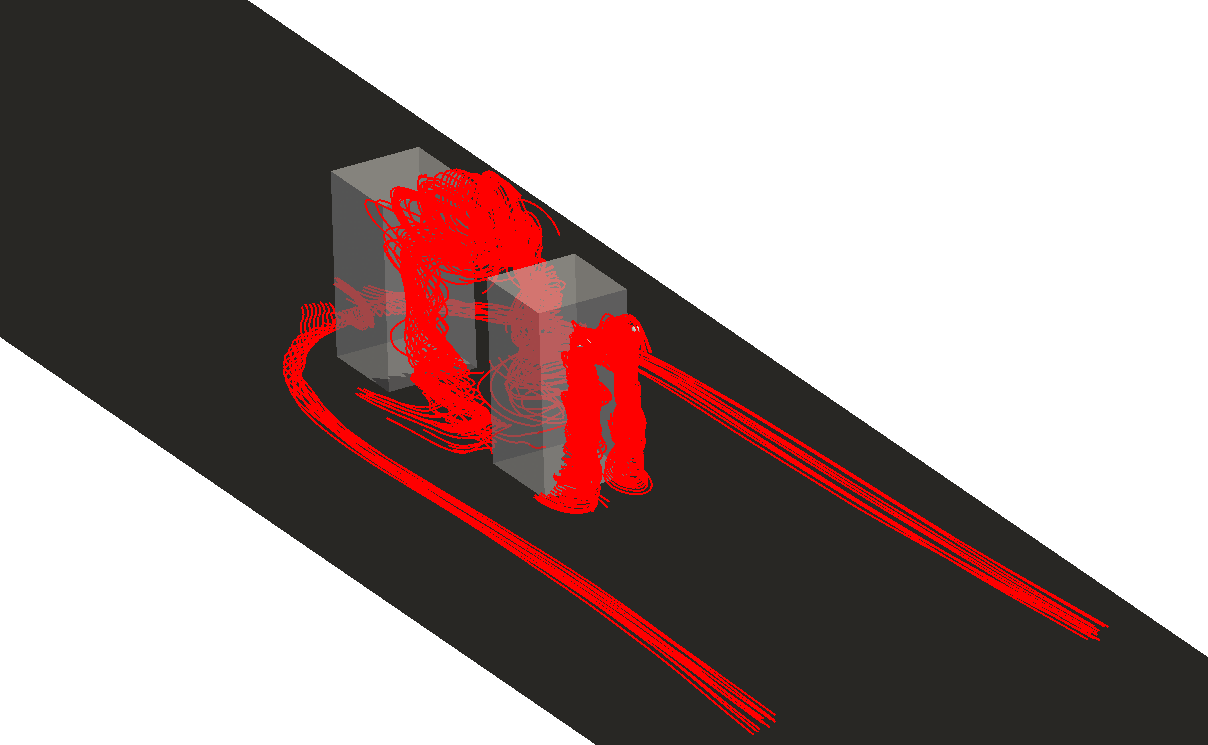}
	\hspace{0.025\textwidth}
         b)
	\includegraphics[trim=150 100 220 100, clip, width=0.26\textwidth]{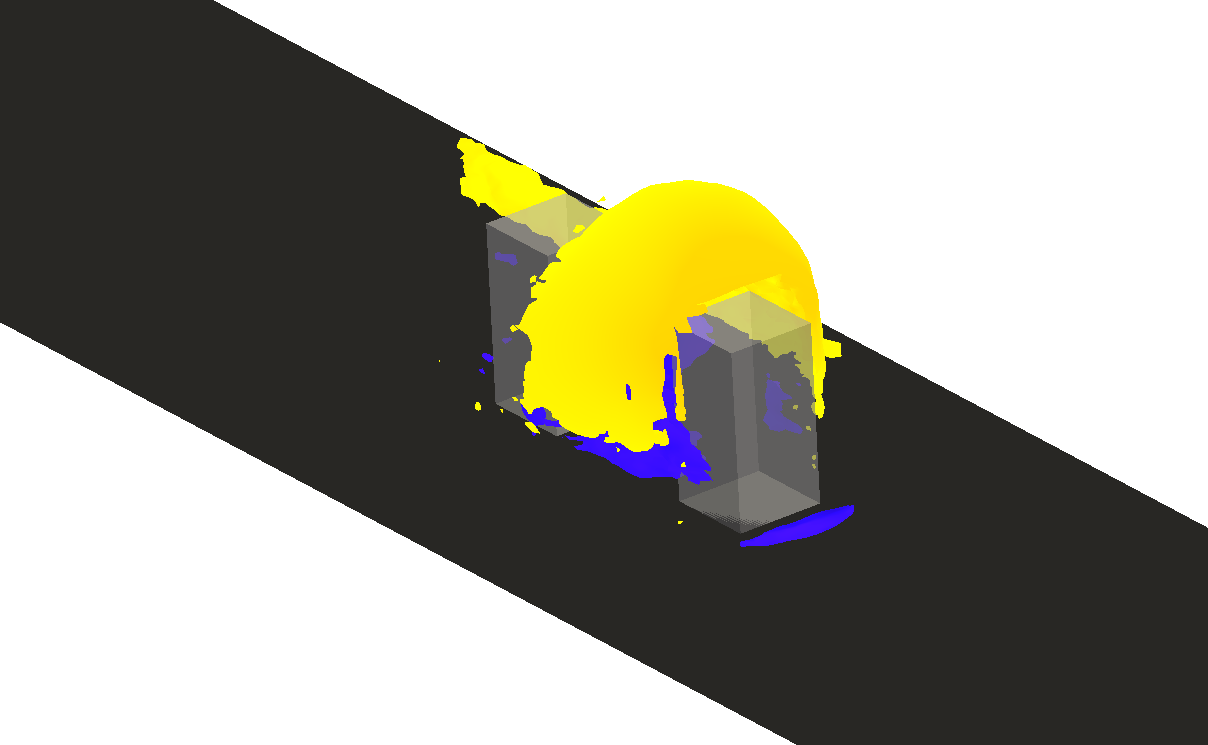}
	\hspace{0.025\textwidth}
        c)
	\includegraphics[trim=150 100 220 100, clip, width=0.26\textwidth]{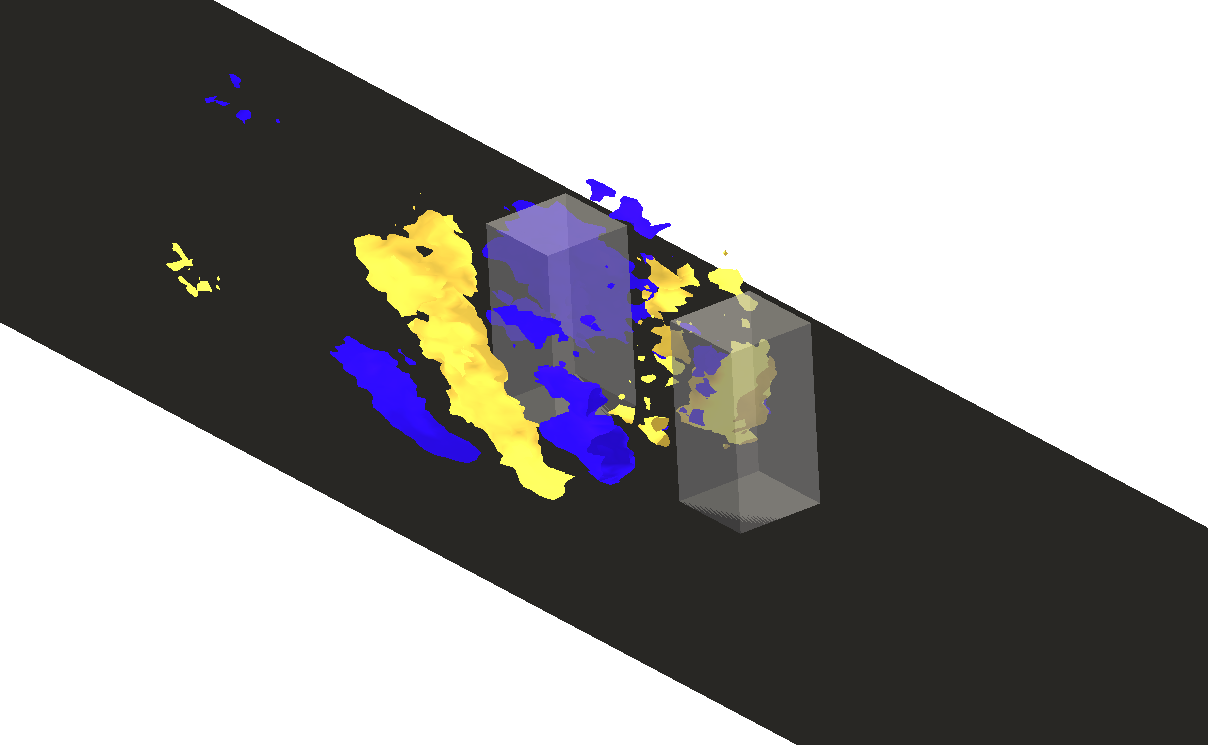}
	\caption{Flow patterns in skimming flow regime in a simplified urban environment. (a) Velocity streamlines representing the arch vortex, (b) and (c) real part of the streamwise velocity of DMD modes: Iso-surfaces normalized with the $L_\infty$-norm for (b) the generator mode ($\omega_m=0.11$), and (c) the breaker mode ($\omega_m=1.1$).  The values used are given by $c_{max}\,\text{max}(U)$ (yellow) and $c_{min}\,\text{min}(U)$ (blue) \label{fig:generator_modes}}
\end{figure}

This first example reviews an application of patterns identification using HODMD that has been published in Ref. \cite{Lazpita}. The problem presented studies a turbulent flow in a simplified urban environment consisting of two buildings, modelled by wall-mounted obstacles, separated by a certain distance. The main goal of this study was to identify the main patterns connected to high concentration of pollution levels. More specifically, it is known that the arch vortex (see Fig. \ref{fig:generator_modes} a), a pattern characteristic of this type of flows which forms dowsntream buildings, plays a crucial role in the dispersion of pollutants in urban areas. HODMD was applied to analyze the flow to identify the main patterns and frequencies leading flow dynamics connected to the presence of the arch vortex. The findings presented in this study are of utmost significance for urban sustainability, as they shed light on the critical factors that contribute to the concentration of pollutants in urban environments. So, we encourage to read the original article Lazpita {\em et al.} \cite{Lazpita} and other related articles for more information \cite{AtzorietalUrban,MartinezetalUrban,CausalityUrban}.

The database for this problem was obtained through numerical simulation applied to a simplified urban environment consisting of two wall-mounted obstacles, where the distance between buildings was varied to obtain different regimes. According to Oke \cite{Oke}, these regimes are called skimming flow, wake interference, and isolated roughness, ordered from shortest to longest distance between buildings. In the case of skimming flow, the spatial dimension of the computational domain is defined for the streamwise, normal and spanwise direction in the interval $x\in [-1,5]$, $y\in [0,2]$ and $z\in [-1.5,1.5]$. respectively. The temporal interval between snapshots is $\Delta t=0.35$. The database used to obtain the results is composed of a spatial grid formed by $100\times 125 \times 50$ points (streamwise, normal and spanwise directions, respectively), and $224$ snapshots. Therefore, the spatial dimension of the problem is $J=1,875,000$, and the temporal dimension is $K=224$. 

HODMD was applied to analyse this database using the parameters listed in Table \ref{tab:calibration_hodmd}, selected based on the criteria discussed earlier.
\begin{table}[H]
    \centering
    \begin{tabular}{|l  l  l |}
    \hline
    \rowcolor{Gray}
    \hline
    \textbf{Parameter} & \textbf{Symbol} & \textbf{Value}
    \\ \hline \hline
    Number of windows &   $d$                 & 50       
    \\
    SVD tolerance     &   $\varepsilon_{svd}$ & $10^{-3}$
    \\
    DMD tolerance     &   $\varepsilon_{a}$   & $10^{-3}$
    \\ \hline
\end{tabular}
\caption{Values of the parameters to choose on the HODMD algorithm for the database presented in this section. \label{tab:calibration_hodmd}}
\end{table}

From this analysis, two types of modes were identified: the so-called {\it generator modes} associated with low frequencies in the computed DMD mode spectrum, which were connected to the presence of the main vortical structures found in the flow (i.e., arch vortex \cite{Monnier,Sumer}), and the so-called {\it breaker modes} that were high frequency modes and were conencted to the wake formed behind the buildings, connected to flow dispersion. These two types of modes are presented in Fig. \ref{fig:generator_modes} b and c, showing the streamwise velocity component. The remaining modes forming the spectrum are combination (formed by non-linear interaction) of these two types of modes.  
%
%

\subsection{Module 1 - application 2: data reconstruction}
SVD algorithm has been applied to repair and enhance the resolution of images complex flows (Sec.~\ref{sec:gappyRepairingGeneralSec}). These two applications are detailed below.

\subsubsection{Application 2.1: data repairing\label{sec:gappyRES}}

Gappy SVD (and HOSVD for databases in tensor form), as introduced in Section \ref{sec:gappy}, depends on two main parameters: the type of initial reconstruction of the database to fill the gaps generally identified with {\it NaN} values (Step 1 of the algorithm), and the number of $P'$ modes that are retained after applying SVD to the reconstructed database (Step 2). On the one hand, in the initial reconstruction, the gaps can be replaced by (i) $0$ values, (ii) the mean value between consecute points or (iii) using a linear (or non-linear) interpolation exploiting the information of the surrounding points. On the other hand, regarding the number of SVD modes $P'$, which are used in the reconstruction of the database (Step 3), using a large number could increase the reconstruction error, since some of these modes could contain  information related to noise or spatial redundancies. On the contrary, using a small number of them could imply loosing relevant information related to the dynamics of the system. Hence, calibration is crucial to ensure the proper performance of the method.

Gappy SVD has been applied to repair a numerical database modelling the two-dimensional wake past a circular cylinder at Reynolds number (compued with the cylinder diameter) $100$. Details about the numerical simulations and the generation of this database can be found in Ref. \cite{VegaLeClaincheBook20}.  The spatial dimension of the computational domain is defined for the streamwise direction in the interval $x\in [-1, 8]d$ and for the normal direction in the interval $y\in [-2, 2]d$, being d the diameter of the cylinder. The dataset analysed is composed by the two velocity components, $449$ points in the streamwise direction and $199$ in the normal direction and $150$ snapshots, equi-distant with time interval $\Delta t=0.2$. Therefore, the data can be organized into a four-order tensor with dimension $2 \times 449 \times 199 \times 100$. From this database, the $\sim 63\%$ of the values are selected randomly and are removed to obtain gaps.

\begin{figure}
    \centering
      \includegraphics[trim=0 100 0 50, clip, width=0.8\textwidth]{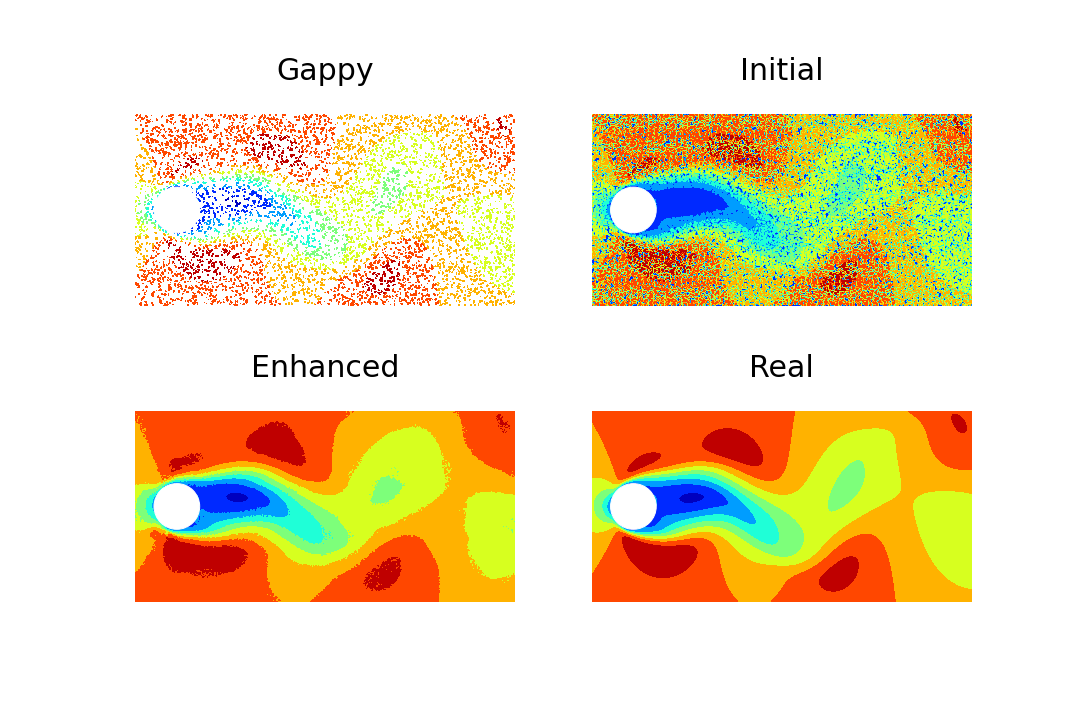} 
    \caption{Gappy SVD algortihm to reconstruct the two-dimensional wake behind a circular cylinder. NaN is found in the $63$\% of the points forming the initial database. From left to right and top to bottom: the corrupted snapshot,  initial reconstruction, final reconstruction and real database.}
    \label{fig:Results_gappy}
\end{figure}

%
%
Gappy SVD is applied using $0$ as initial reconstruction value (replaicing Nan), and the number of retained SVD modes selected is $10$. Figure \ref{fig:Results_gappy} shows the initial database, the inital reconstruction carried out by the algorithm (iteration i=1) and compares the real solution with the final reconstruction. The RRMSE eq. (\ref{eq:rrmse}) of the reconstruction is smaller than $2$\%.
%

For reconstruction of datasets considering the temporal component, similar results are obtained, using the algorithm Gappy HOSVD instead. See details in Ref. \cite{ModelFLOWsappWeb}.

\subsubsection{Application 2.2: superresolution}

This second type of application of SVD algorithm is tested to enhance the resolution of a database. The only parameter necessary to set is the final resolution that is desired to obtain in the super-resolved database (see details in Section \ref{sec:superResSVD}). As described in the methodology, the resolution of the database is doubled in each iteration the algorithm is applied, so the final resolution will be a number that is calculated as the power o $2$ times the initial resolution.  

As in the previous section, the method is tested to enhance the resolution of a database modelling the two-dimensional wake behind a circular cyinder at Reynolds number $100$. The dimension of the database, organized in tensor form,  is $2 \times 449 \times 199 \times 150$, corresponding to the velocity components, grid points along the streamwise and normal directions, respectively, and the number of snapshots. For this application, the dimension of the database is reduced to $2 \times 63 \times 63 \times 100$, so the spatial dimensions have been downsampled by a factor or $2^3=8$ (only $1$ every $8$ points are reained in the reduced database. In ModelFLOWs-app then it would be necessary to include the downsampled database and the parameter $3$, to enhance the resolution to the original dimension.



\begin{figure}
    \centering
      \includegraphics[width=\textwidth]{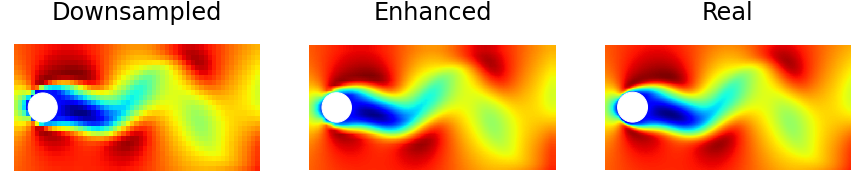} 
4    \caption{Application of the superresolution algorithm to a database extracted from the flow past a cylinder. From left to right: the downsampled snapshot snapshot, the solution gave by the algorithm and the real snapshot extracted from the simulation.}
    \label{fig:Results_resolution}
\end{figure}

 Figure \ref{fig:Results_resolution} shows the downsampled database and compares the resolution enhanced database with the original one, where the RRMSE eq. (\ref{eq:rrmse}) computed is smaller than $3\%$.

For reconstruction of datasets considering the temporal component, similar results are obtained, using the algorithm HOSVD instead. See details in Ref. \cite{ModelFLOWsappWeb}.

\subsection{Module 1 - application 3: predictive models}

Let us examine a specific data-oriented ROM developed using HODMD (Sec.~\ref{sec:HODMD})  for temporal prediction. This example is presented in detail in Refs. \cite{LeClaincheVegaPoF17,VegaLeClaincheBook20} .The data-driven predictive ROM is developed to expedite the computational efficiency of Nek5000 \cite{Nek5000}, an open source spectral element code employed for solving the incompressible continuity and Navier-Stokes equations \cite{Batchelor}, thereby enabling the computation of the system's final attractor for temporal forecasting. Our focus lies on the three-dimensional wake of a circular cylinder in the context of incompressible fluid dynamics, a well-established benchmark problem \cite{Williamson89,Williamson96}. For this problem, we define the Reynolds number (with the cylnder diameter) as $210$. The spanwise length of the computational  geometry studied corresponds to a spanwise wavenumber of $4$. Descriptions of the computational domain, the mesh, and the analysis of the accuracy  of the numerical simulations can be found in \cite{LeClaincheVegaPoF17}. To construct the data-driven ROM, we firt perform numerical simulations and then we use a collection of $500$ snapshots, equi-distant with time interval $\Delta t=1$, obtained from the saturated regime of such simulations, specifically within the time interval $575\leq t\leq 825$. These snapshots serve as the basis for developing the data-driven model.

\begin{table}[H]
    \centering
    \begin{tabular}{|l  l  l |}
    \hline
    \rowcolor{Gray}
    \hline
    \textbf{Parameter}   & \textbf{Symbol}       & \textbf{Value}
    \\ \hline \hline
    Number of windows    &   $d$                 & 250 \\
    SVD tolerance        &   $\varepsilon_{svd}$ & $10^{-4}$\\
    DMD tolerance        &   $\varepsilon_{a}$   & $3\cdot10^{-3}$
    
    \\ \hline
\end{tabular}
\caption{Values of the parameters to choose on the predictive HODMD algorithm for the database presented in this section. \label{tab:calibration_pred_hodmd}}
\end{table}

\begin{figure}[!h]
          \vskip-0.25cm
 \begin{center}
\includegraphics[width=0.45\textwidth]{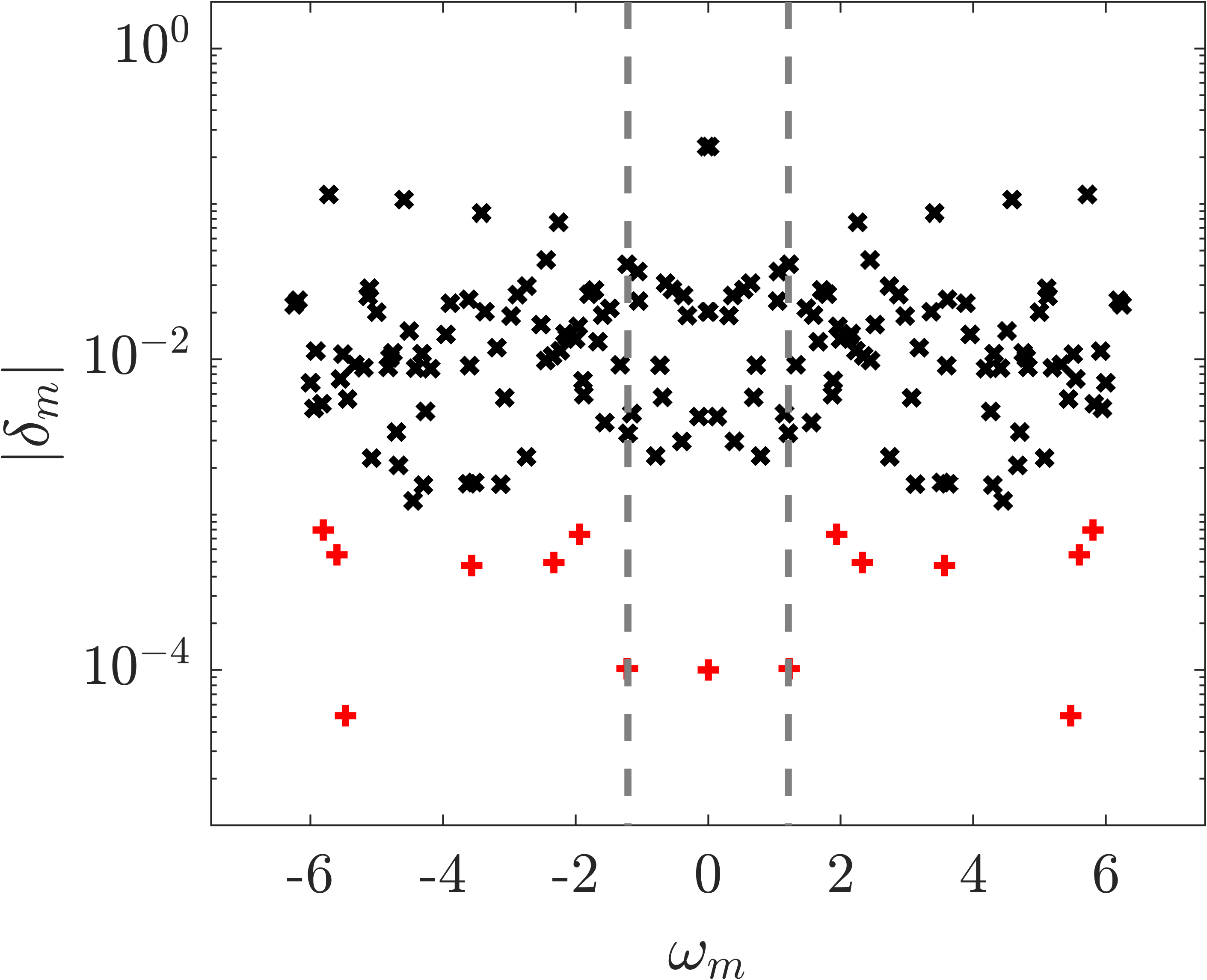}
\includegraphics[width=0.45\textwidth]{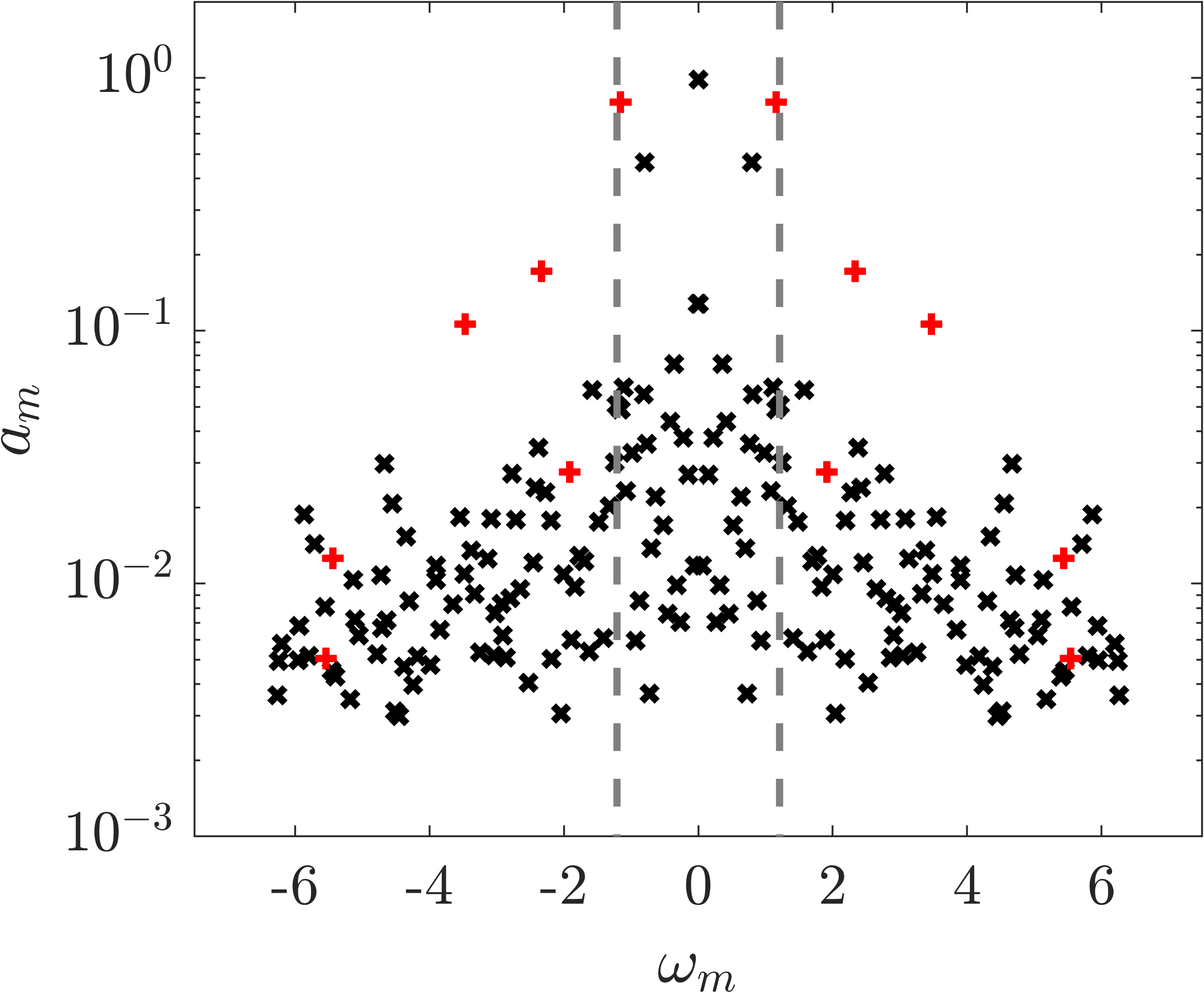}
\end{center}
\vskip-0.75cm
\caption{Frequencies vs. damping rates (left) and mode amplitudes (right) of the DMD modes calculated in the transient of a numerical simulation modelling the three-dimensional wake behind a circular cylinder.  Red: modes with smaller damping rates, retained to form the ROM. Black: remaining modes. Fundamental frequencies are indicated by vertical lines.
\label{fig:spectrum_pred_hodmd}}
 \end{figure}

HODMD is then applied to analyse the previous dataset, where the calibration of the method involved a careful selection of values, presented in  Tab. \ref{tab:calibration_pred_hodmd}, ensuring consistency and robustness in the obtained results \cite{LeClaincheVegaSoria17}. Within our analysis, we have distinguished two distinct types of DMD modes: transient modes characterized by $\delta < 0$ and permanent modes with $\delta \simeq 0$ (see the DMD expansion eq. (\ref{ab00})). 
Visual representations of the relationships between damping rates, mode amplitudes, and retained frequencies can be found in Fig. \ref{fig:spectrum_pred_hodmd}.

To construct the ROM, permanent modes are selected based on a condition for their growth rate: $\vert \delta_m \vert <\varepsilon=10^{-3}$. The HODMD method is then employed to extrapolate the results, excluding transient modes, to the attractor for $t\geq 1900$. For such application, ModelFLOWs-app follows the next steps: (i) the original dataset is reconstructed utilizing the DMD expansion eq. (\ref{ab00}) only using the selected DMD modes, (ii) the growth rate of the selected DMD modes is set to $0$ (although it is also possible not to change the growth rate of the modes, as function of the needs of the ROM), (iii) the temporal term of the DMD expansion is set as $t\geq 1900$. The ROM predicts then the temporal evolution of the flow for temporal instants $\geq 1900$.  

\begin{figure}[h!]
 \vskip-0.25cm
 \begin{center}
  \includegraphics[width=0.8\columnwidth]{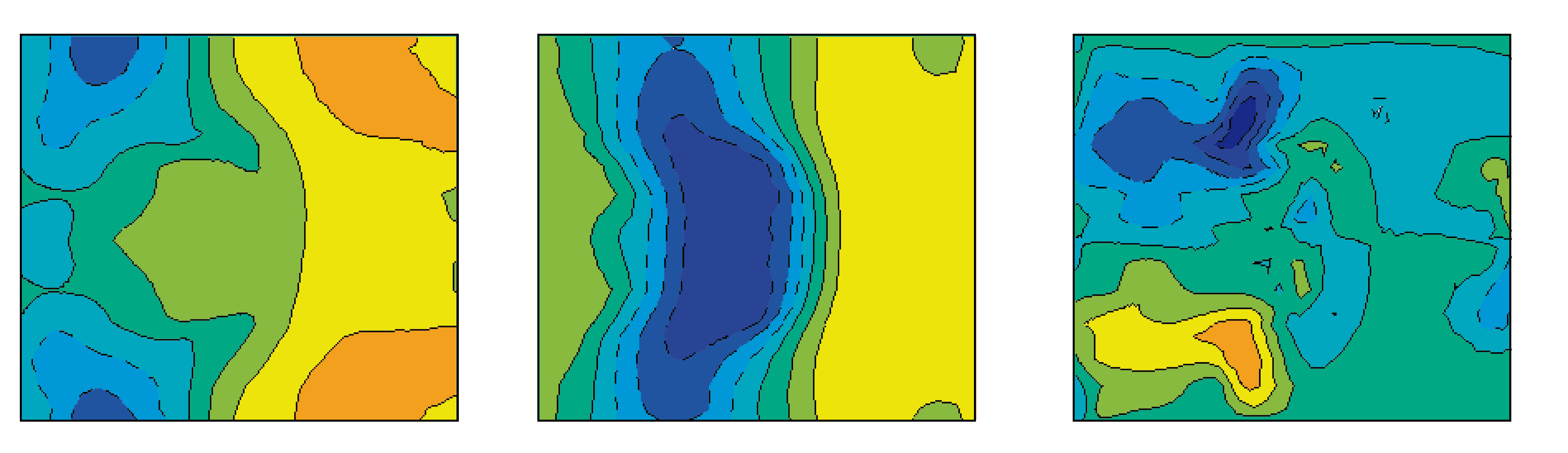}\\
  Original solution\\
  \vskip0.5cm
  \includegraphics[width=0.8\columnwidth]{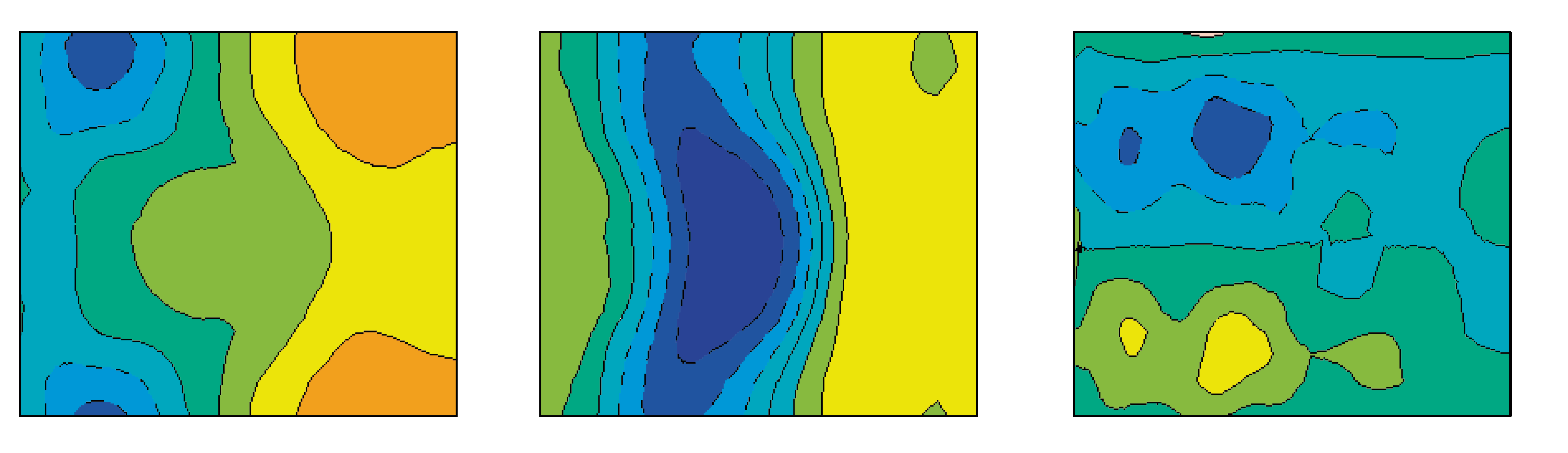}\\
  Data-driven ROM using HODMD\\
    \vskip0.5cm
  \includegraphics[width=0.6\columnwidth]{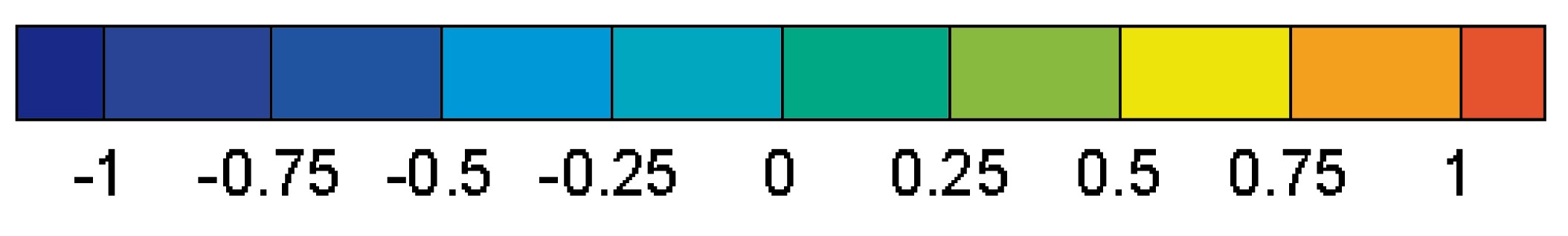}
  \end{center}
  \vskip-0.25cm
\caption{Predictive ROM using HODMD tested in the three-dimensional way of a circular cylinder. Streamwise (left), normal (center), and spanwise (right) velocity components in the mid $x-z$ plane for a representative snapshot of the attractor at $t=2900$. The data has been normalized using the maximum velocity value from the original data. \label{fig:ROMcylinder}}
\end{figure}

The predicted solution exhibits a RRMSE error eq. (\ref{eq:rrmse}) of $\sim 6$\% when compared to the real solution. Figure \ref{fig:ROMcylinder} shows a representative snapshot from the attractor, comparing the outcomes of the data-driven ROM with the original data. Notably, the ROM approximation performs worse in the spanwise velocity component compared to the other two components, mainly due to its significantly smaller magnitude. However, it can be inferred that the spatio-temporal symmetry in the spanwise velocity component is essentially preserved. The seep-up factor of this ROM compared to the numerical simulations is larger than $100$.

\subsection{Module 2 - application 1: patterns identification}

Autoencoders  algorithm (Sec.~\ref{subsec:module2_AEs}) has been applied to identify temporal patterns in complex flows. More specifically, we applied this technique on the flow generated by two planar synthetic jets.
These devices are characterized by the periodic movement of a membrane or piston inside a cavity. Each periodic cycle consists of an injection and suction phases in which the fluid is ejected and reintroduced into the cavity through a jet nozzle, respectively.
The database analysed model two synthetic jets working synchronously at Strouhal number (defined with the jet diameter) $0.03$ and Reynolds number (defined with the jet diameter) $100$ (from Ref. \cite{MunozLeClainche2022}). The spatial dimension of the computational domain is defined for the streamwise and normal direction in the interval $x\in [0.9,35.6]D$ and $y\in [-10,5.7]D$, respectively, where $D$ refers to the diameter of the nozzle and the axis center is the center of upper jet exit. The temporal interval between snapshots is $\Delta t=5.34$e-02 $U/D$, where $U$ is the characteristic velocity of the flow.

The main patters of the flow are analysed using POD and HODMD methods (both from ModelFLOWs-app module 1), and this section shows a new application where also AEs are used to extract the main patterns connected to the flow dynamics. More details about the results presented in this section can be found in Ref.  \cite{Munozetal2023_preprint}.

\begin{figure}
    \centering
      \subfloat[HODMD] {\includegraphics[height=0.10\textheight]{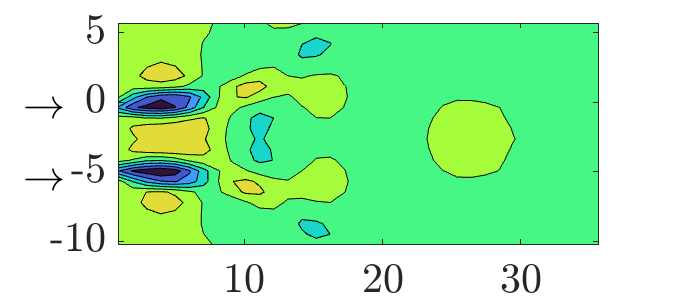} \label{subfig:AEs_DMD1}} 
      \subfloat[AEs] {\includegraphics[height=0.10\textheight]{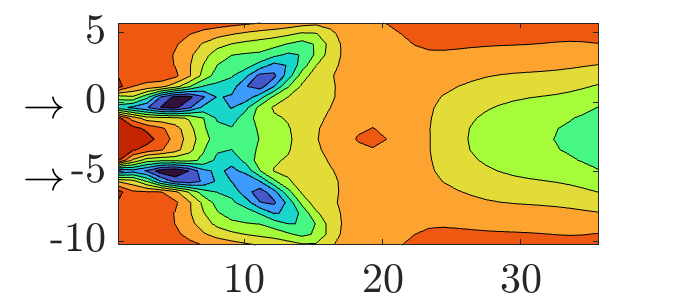}   \label{subfig:AEs_AEs1}} 
      \subfloat[POD] {\includegraphics[height=0.10\textheight]{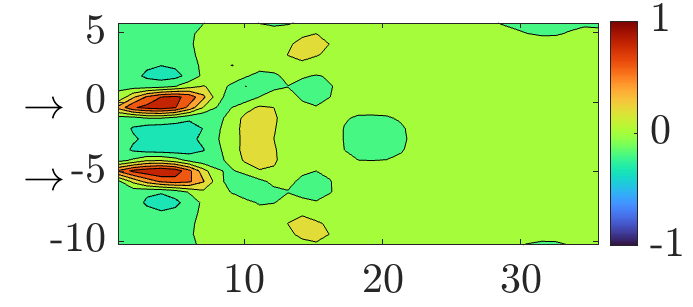}   \label{subfig:AEs_POD1}} 
    \caption{a) AEs, b) HODMD and c) POD first contributing mode on two synthetic jets. Arrows represent the jets' input. Legend from c) applies to all pictures.}
    \label{fig:Results_AEs}
\end{figure}

Figure~\ref{fig:Results_AEs} compares the dominant mode identified by HODMD, AEs and POD methods. In previous research \cite{MunozLeClainche2022}, HODMD relates this mode with the oscillation frequency driving the synthetic jet, Strouhal number $0.03$. In the results obtained, these three modes present some differences, as expected since we use three different methodologies. Nevertheless, the high intensity areas found in the three modes are similar: the three modes recognize two high intensity regions in form of two ovals at the jets exit. A region with still high (but less than before) intensity  extends further downstream the two jet exits. This region is more strengthned in AEs. The difference found in the shape of the modes could be connected to the non-orthogonality behind AEs calculations. More results and a detailed explanation can be found in Ref. \cite{Munozetal2023_preprint}. Also Ref. \cite{ae_modal} shows a detailed comparison of differet type of AEs and SVD modes, as well as their properties for reduced order modelling in turbulent urban flows.

The calibration of AEs depends on the characteristics of the database analyzed and its spatial and temporal dimensions, $J$ and $K$, respectively. The parameters to choose in ModelFLOWs-app are the training percentage, $\%_{train}$, that splits the data into the training and test size ($K_{training}=\%_{train} K$, $K_{test}=(1-\%_{train}) K$, respectively)  and it is recommended to be smaller or equal to $80$\%; the batch size, $N_{batch} \in [1, K_{training}]$ is the size of the packages used for the training, and it is recommended to be a power of 2, being usually 32 and 64 the best values\citep{bengio2012practical}; the number of AE modes retained, $M <  K_{training}$, also called as encoding dimension; and the maximum number of epochs, $N_{epochs}$, is the maximum number of passes the algorithm takes around the training data, and possible examples of its value are 100, 200, 500,$\cdots$.  We propose high numbers for this parameter because we have implemented early stopping in the code. Therefore, if $N_{epochs}$ is high, the early stopping will stop the training when the convergence is achieved, and it is too low, the training could stop before without assuring convergence. Finally, it is notorious that the number of AE modes retained is the parameter that will most affect the results. If the main application of AEs is to observe the main dynamics of the flow, we advise retaining a small number (i.e., 5, 10, 20), but to obtain small reconstruction errors, the number of modes retained should be larger.  Based on these recommendations, the parameters chosen in the  database analysed in this section, with $J$= 1980 and $K$=4369, are stored in Tab.~\ref{tab:calibration_AEs}.
\begin{table}[H]
    \centering
    \begin{tabular}{|l  l  l |}
    \hline
    \rowcolor{Gray}
    \hline
    \textbf{Parameter} & \textbf{Symbol} & \textbf{Value}
    \\ \hline \hline
        Training percentage             &   $\%_{train}$    & 80 \%     \\
        Batch size                      &   $N_{batch}$     & 32        \\
        $\#$ modes (autoencoder)   &   $M$             & 10        \\
        Epoch number                    &   $N_{epochs}$     & 200    \\ \hline
    \end{tabular}
   \caption{Calibration in AEs algorithm for patters indentification in two plana synthetic jets.  Database with dimension $J$= 1980 and $K$=4369.}
    \label{tab:calibration_AEs}
\end{table}

\subsection{Module 2 - application 2: data reconstruction}

Hybrid ROMs combining SVD with deep learning architectures (Sec.~\ref{sec:reconstructionDNN}) has been applied to reconstruct downsampled databases in complex flows. 
This application is explained in detail in Ref. \cite{paperPaula}, where the authors propose a hybrid model, combination of SVD with neural networks. The neural network considered consists on combining two autoencoders that work in parallel and are joined into the last layer. The architecture has been proved to be robust and generalizable, wich allows reconstructing databases with a low reconstruction error, from databases containint information from a few sensors (as it is the case in experimental measurements).

As in the example presented in module 1 (Sec.~\ref{sec:gappyRES}), the case analysed is the saturated regime modelling the two-dimensional wake behind a circular cylinder at Reynolds number (defined with the cylinder diameter) $100$. The database analysed has been obtained numerically, where details about the performanc of the numerical simulation can be found in Ref. \cite{VegaLeClaincheBook20}. The spatial dimension of the computational domain is defined for the streamwise direction in the interval $x\in [-1, 8]d$ and for the normal direction in the interval $y\in [-2, 2]d$, being d the diameter of the cylinder. The database analyised is formed by $3$ variables, the two velocity components (streamwise and normal) and the spanwise vorticity component, $449$ spatial points along the streamwise direction and $199$ points along the normal direction, and $150$ snapshots equispaced in time with an interval of $0.2$. From this database, a downsampled tensor is created, taking one every $30$ points in both spatial directions, resulting in a tensor with dimensions $3 \times 15 \times 7 \times 150$. 

\begin{figure}
    \centering
    {\includegraphics[width = 1.1\textwidth]{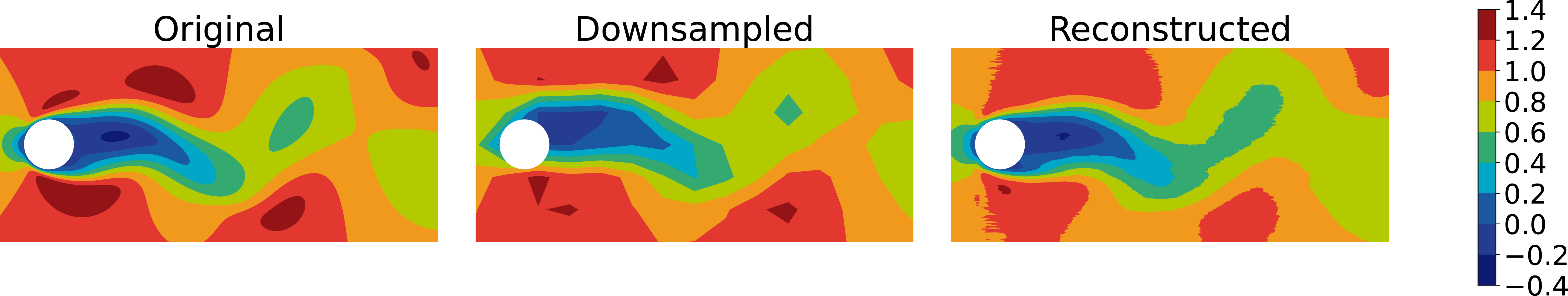}} \\
    {\includegraphics[width = 1.1\textwidth]{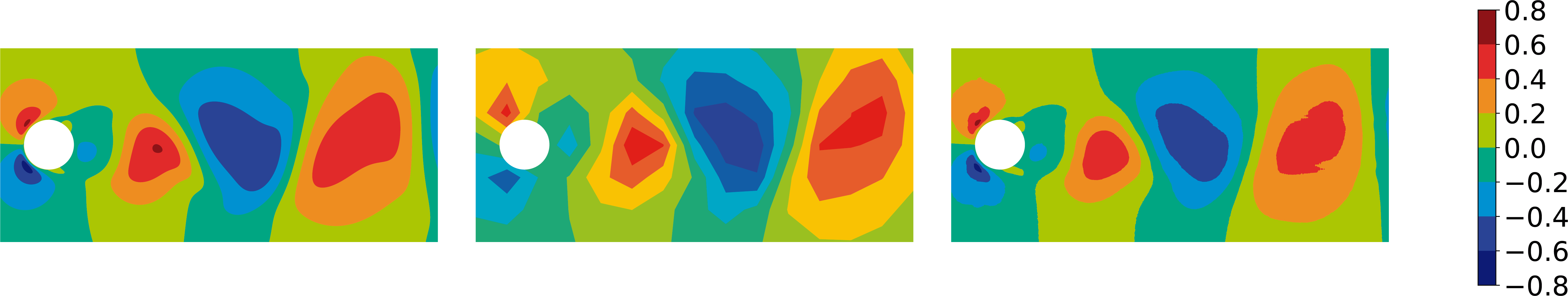}} 
    \caption{Reconstruction of a representative snapshot of the flow past a cylinder using ModelFLOWs-app. From left to right: original snapshot, the downsampled snapshot (input of the neural network) and the reconstructed snapshot (output of the algorithm). From top to bottom: Streamwise and normal velocity components.}
    \label{fig:Results_NNreconst}
\end{figure}

Figure \ref{fig:Results_NNreconst} shows the downsampled database and compares the reconstruction of a representative snapshot of the database, carried out using the present methodology, with the original solution. The RRMSE eq. (\ref{eq:rrmse}) in the reconstruction is smaller than  $8\%$ in the whole database analysed. The algorithm has also been successfully tested with initial databases composed by a very small number of grid points (i.e. $8$ in each spatial direction). More details can be found in in Ref. \cite{paperPaula}.
%

The calibration of the method depends on several parameters, which are listed in Tab. \ref{tab:calibration_NNReconst}. If selected, the database can be pre-processed with the parameter $1^{st} Scaling$, as well as the matrices that will enter the deep learning model, in that case the parameter is called as $2^{nd} Scaling$. The parameter $\%_{train}$ refers to the data used in the training process, which is recommended to be less than the $80\%$ of the samples. The batch size $N_{batch}$ can be also selected, as in the previous section, as well as the number of epoch ($N_{epoch}$).

For the deep learning model, ModelFLOWs-app offers the possibility to selecting and using the optimal model hyperparameters of the deep learning architecture: we use {\it RandomSearch Keras} tuner.  Additionally, these parameters can be directly selected by the user. These are: (i) the number of neurons $N_{neurons}$, the activation function $AF$, the loss function $l_f$ and the learning rate $l_r$. 
\begin{table}[H]
    \centering
    \begin{tabular}{|l  l  l |}
    \hline
    \rowcolor{Gray}
    \hline
    \textbf{Parameter} & \textbf{Symbol} & \textbf{Value}
    \\ \hline \hline
        First Scaling             &   $1^{st} Scaling$    & $No$     \\
        Second Scaling             &   $2^{nd} Scaling$    & $No$     \\
        Training size                        &   $\%_{test}$     &   0.8     \\
        Batch size                      &   $N_{batch}$     & 23        \\
        Epoch number                    &   $N_{epochs}$     & 500    \\
        Hyper parameterization              &   $Hyper$         &   No     \\
        Activation functions (only for $Hyper$ = No) &   $AF$      &   ReLU \\
        \# Neurons (only for $Hyper$ = No)  &   $N_{neurons}$       &  13     \\  
        Learning rate (only for $Hyper$ = No)  &   $l_r$            &   0.002   \\
        Loss function (only for $Hyper$ = No)  &   $l_f$            &  $mse$ \\
        \hline
    \end{tabular}
   \caption{Parameters to choose on the data reconstruction algorithm and the values selected for the database presented in this section.}
    \label{tab:calibration_NNReconst}
\end{table}

\subsection{Module 2 - application 3: time-series forecasting models
\label{sec:ResPredictionNN}}

Time-series forecasting models based on deep learning (Sec.~\ref{sec:predictionDNN})  have been applied to two diferent problems modelling different types of complex flows. The main goal of this application is to extract data from the transient region of a numerical simulation and to predict the evolution of the simulation, so these ROMs have been developed to speed-up numerical simulations. 

To show the good properties and robustness of the models proposed, two different methodologies have been followed. We first apply the hybrid framework (HybridDL) presented in Refs.  \cite{AbadiaetalExpSystAppl22,paperAdricombustion}, where the proposed models are a combination of SVD and deep learning architectures. Secondly we follow a fully deep leaning framework (FullyDL), as in Ref. \cite{MataLeon2023}, where the proposed models are fully based on deep learning architectures. It is worth to mention that for this second application, in Ref. \cite{MataLeon2023}, dimensionality reduction was also carried out by HODMD, instead of SVD.  Nevertheless, the present article only shows the application of the deep learning architecture itself. As seen in this section, the two methodologies available in ModelFLOWs-app to develop predictive ROMs can be applied to different kind of problems.

\begin{figure}[H]
	\centering
	\includegraphics[width=0.65\textwidth]{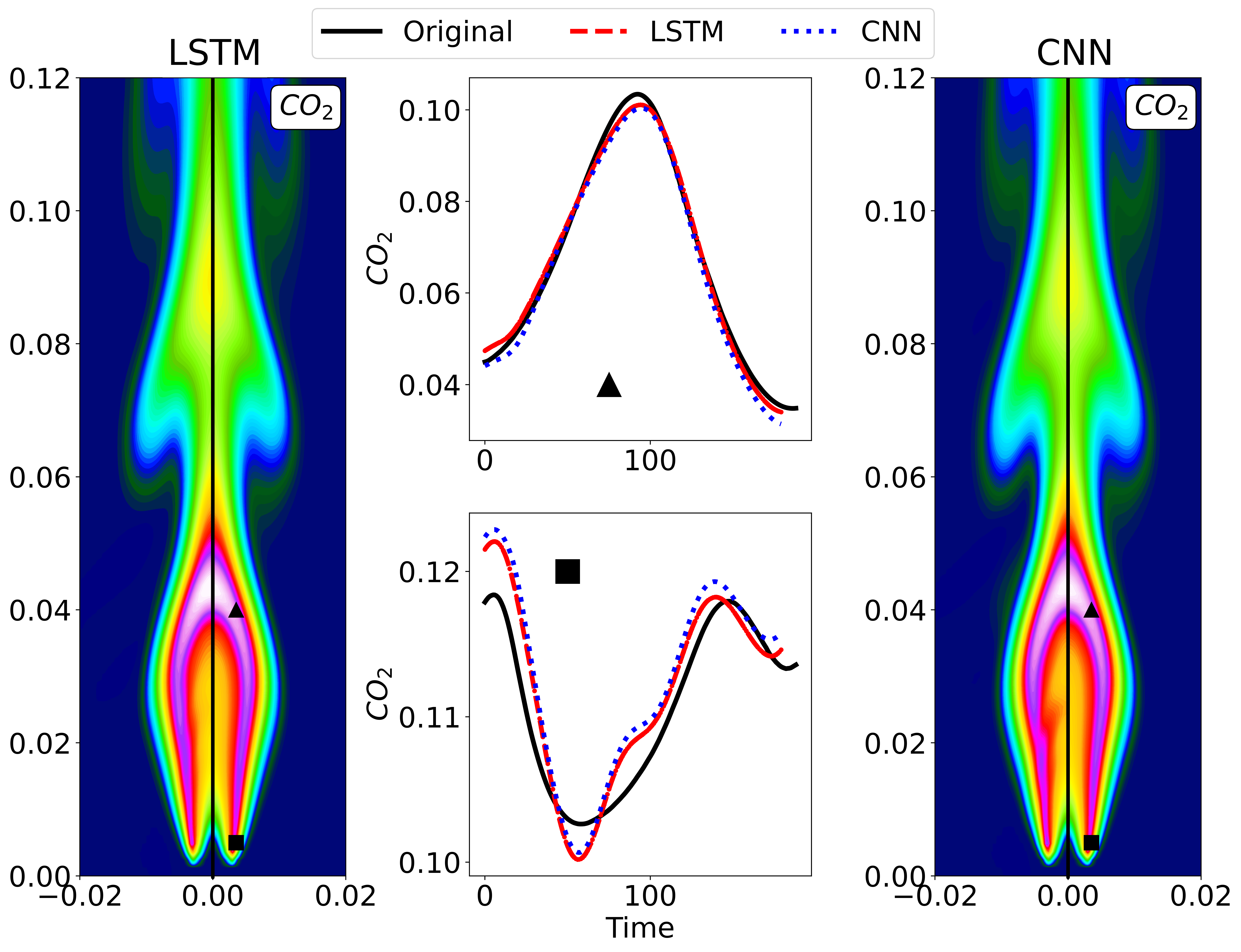}
	\caption{Predictive hybrid ROM to predict reactive flows. Representative snapshot showing concentration of $CO_2$. Predictions carried out using RNN (left) and CNN (right) architectures. Right and left parts of each contour show the original snapshot the prediction. Mid par of the figure: temporal evolution of temperature  at two characteristic points of the computational domain. Points extracted where the triangle and square are located in the figure contour.}
	\label{fig:Figure10}
\end{figure}

As mentioned before, HybridDL models first apply SVD to the database. This pre-processing allows to reduce the number of trainable parameters required by the models, simplifying the training. But, at the cost of loosing part of the spatial information, since the spatial dimension tensor is flattened into a single vector before applying SVD, forming the snapshot matrix eq. (\ref{ab0}). This methodology shows good performance in several examples: the three-dimensional wake of a circular cylinder, synthetic jets \cite{AbadiaetalExpSystAppl22}, and reactive flows (axisymmetric, time varying, non-premixed, co-flow nitrogen-diluted methane flame \cite{paperAdricombustion}).  Results about the latter case are shown. More information about the numerical simulacion of the laminar flame and the numerical set-up can be found in Refs. \cite{d2020impact, d2020adaptive}. From this detailed simulation, a database is extracted, composed by $10$ variables (the temperature and 9 chemical species). The spatial dimension of the computational domain is defined for the axial direction: $100$ points in the interval $z\in [0, 0.12]$; and for the radial direction: $75$ points in the interval $r\in [0, 0.02]$, respectively.
The number of snapshots extracted is $K = 999$, equidistant in time with $\Delta t = 2.5 \times 10^{-4} $. Figure \ref{fig:Figure10} shows the reactive flow test case. It compares the real solution with the predictions carried out using RNN and CNN architectures. The number of SVD modes retained in the first step of the model is $18$. The figure shows a representative snapshot of the $CO_2$ mass fraction and the evolution in time of temperature field extracted at two representative points from the computational domain. The RRMSE eq. (\ref{eq:rrmse}) in the predictions is smaller than $4 \%$ and $3\%$ for the CNN and RNN architectures, respectively, and the speed-up factor in the numerical simulations is higher than $100$ in both cases. See more details about this example in Ref. \cite{paperAdricombustion}.

\begin{figure}[H]
    \centering
    \includegraphics[width=1\textwidth, angle=0]{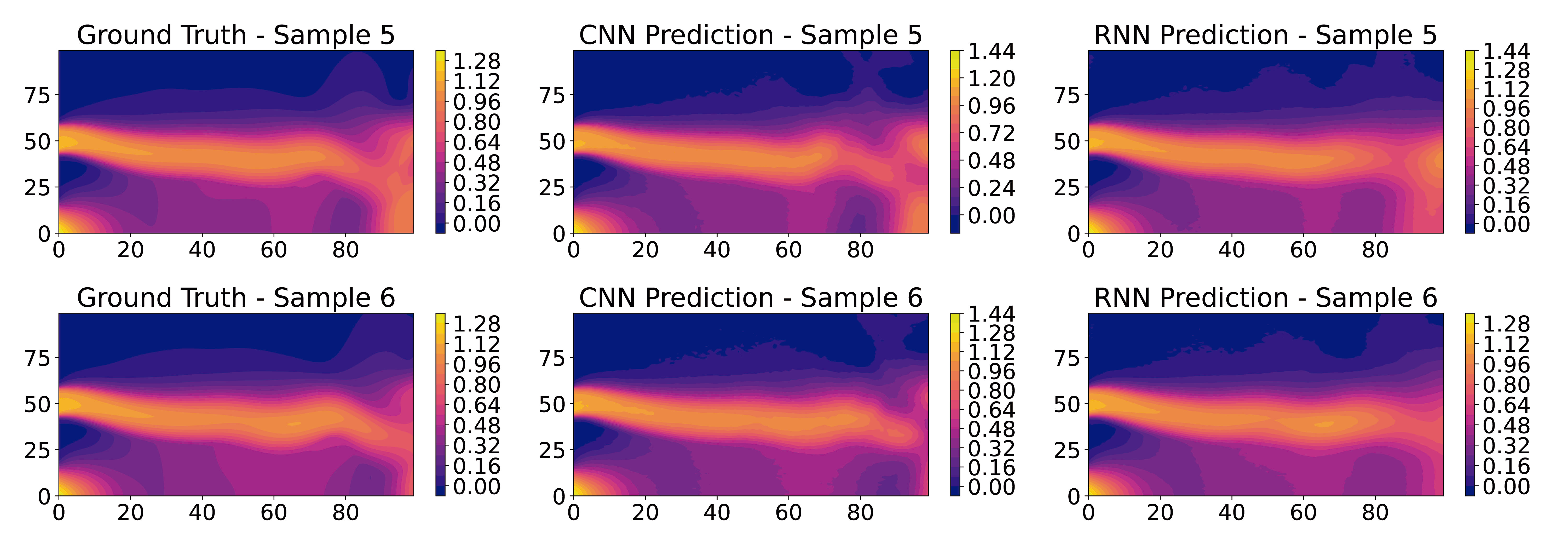}
    \caption{Predictions carried out using models insisde FullyDL framework. Streamwise velocity in the original solution (left), and in the predictions using CNN (middle) and RNN (right) architectures.}
    \label{fig:results_fullydl_rnn}
\end{figure}

The second example uses the methodology FullyDL, which in contrast to HybridDL, any dimensionality reduction is carried out in the original database. This increases the number of trainable parameters and, therefore, the complexity of the neural network training. However, we keep the spatial information contained in the snapshots, i.e., the flow structures that define the flow dynamics. This methodology has been successfully tested in several problems modelling complex flows, as explained in Refs.  \cite{LopezMartinetal2020,MataLeon2023}. This section presents the results of applying these models to predict the evolution of a two-phase flow concentric jet. The flow comprises two liquid jets, consisting of two incompressible, viscous, and immiscible fluids, with Reynolds numbers of $Re_{1} = 30$ and $Re_{2} = 200$ for each phase, respectively, and a Weber number of $We = 80$. Both jets arise from two nozzles separated by a gap, whose length can be close to zero.  The spatial dimension of the computational domain is defined for the streamwise direction in the interval $x\in [0, 16]$ and $y\in [0, 8]$, respectively. The temporal interval between snapshots is $\Delta t=0.005$. Details about the numerical simulations carried out to solve this problem can be found in Ref. \cite{MataLeon2023}.  The database used to obtain the results in Figure \ref{fig:results_fullydl_rnn} is composed of a spatial grid formed by $100\times 100$ points for the streamwise velocity, and $301$ snapshots. Therefore, in this case the spatial dimension of the problem is $J=100\times 100$ (in this framework we do not flatten the spatial dimension), and the temporal dimension is $K=301$. Figure \ref{fig:results_fullydl_rnn} compares two representative snapshot of the predictions carried out using the CNN and RNN models. The RRMSE eq. (\ref{eq:rrmse}) computed in these predictions is $0.044$ and $0.1$ for the CNN and RNN architectures, respectively. The speed-up in the numerical simulations is $9.88$ and $9.03$ for the CNN and RNN architectures, respectively.

In each methodology, we developed two different models: CNN and RNN, which is based on the Long Short-Term Memory (LSTM) architecture \cite{Hochreiter1997LSTM}. It is important to note that the LSTM architecture only uses a vector structure, which is not a problem in HybridDL because the spatial dimension was already flattened to apply SVD. However, the FullyDL framework takes the original snapshots as input, which means that the spatial dimension needs to be flattened. Due to this, the RNN model exhibits significantly better performance in the HybridDL framework compared to FullyDL. On the other hand, the CNN model shows better performance in the FullyDL framework than in HybridDL. This is because Convolutional Neural Networks were developed for analyzing snapshots, particularly in the field of computer vision\cite{Lecun1998CNN}. Therefore, this architecture is capable of better capturing the spatial structures present in the snapshots.

As seen in Tabs. \ref{tab:calibration_HybridDL}, \ref{tab:calibration_FullyDL} the calibration of the  RNN and CNN architectures differ depending on the framework: HybridDL or FullyDL. Details are presented below.


\subsubsection{Calibration of models in HybridDL framework}

\begin{table}[H]
    \centering
    \begin{tabular}{|l  l  l |}
    \hline
    \rowcolor{Gray}
    \hline
    \textbf{Parameter} & \textbf{Symbol} & \textbf{Value}
    \\ \hline \hline
        \# modes (SVD)                      &   $N$             &   18      \\
        First scaling                     &   $1^{st}Scaling$               &   Auto  \\
        Second scaling                     &   $2^{nd}Scaling$               &   MpM      \\
        Testing size                        &   $\%_{test}$     &   0.2     \\
        Validation size                     &   $\%_{val}$      &   0.15     \\
        Batch size                          &   $N_{batch}$     &   12       \\
        \# epochs                           &   $N_{epochs}$    &   400      \\
        \# input samples                    &   $k$             &   10      \\
        \# time ahead predictions           &   $p$                 &   6       \\
        Hyper parameterization              &   $Hyper$         &   No     \\
        Hidden activation functions (only for $Hyper$ = No) &   $HiddenAF$      &   ELU \\
        Output activation functions (only for $Hyper$ = No) &   $OutAF$      &   Tanh \\
        \# Neurons (only for $Hyper$ = No)  &   $N_{neurons}$       &   100     \\
        Shared dimension (only for $Hyper$ = No)   &   $SharedDim$    &   80      \\
        Learning rate (only for $Hyper$ = No)  &   $l_r$            &   0.005   \\
        Loss function (only for $Hyper$ = No)  &   $l_f$            &   custom\_loss \\
        \hline
    \end{tabular}
    \caption{Hyperparameters of models belonging to the HybridDL framework. The symbol $\#$ means number and the Value column shows the configuration used in Ref. \cite{paperAdricombustion}.}
    \label{tab:calibration_HybridDL}
\end{table}

Both RNN and CNN models share the same hyperparamenters for calibration, listed in Table \ref{tab:calibration_HybridDL}. Note this calibration depends on the database. From the mentioned case, the reactive flow, presented in Ref. \cite{paperAdricombustion}, the database analysed is formed by: the temperature and the most important $9$ chemical species, $100$ points on the streamwise direction, $75$ on the normal direction and $999$ snapshots. This gives a fourth-order tensor of dimensions $10 \times 100 \times 75 \times 999$.

The hyperparameter $1^{st} Scaling$ indicates the general scaling to be applied to the database. This step is necessary in combustion databases, as the variables can differ significantly in magnitude among them. In this case, $auto scaling$ has been used, which utilizes the standard deviation of each variable, ensuring that all variables have equal importance \cite{PCA}. After this scaling step, the tensor is reshaped into a matrix, grouping all dimensions except the temporal one into a single dimension. This results in a matrix with dimensions of $75000 \times 999$. Subsequently, SVD is applied, and the first $N$ modes are retained, which contain the large scales.

The term $2^{nd} Scaling$ refers to the scaling applied to the temporal coefficients of the SVD, specifically for the reduced database ($\widehat{\bV}_1^K$) used for training the models. In our case, $MpM$ scaling has been employed, as it has been demonstrated to be suitable when dealing with non-periodic databases \cite{paperAdricombustion}.

The hyperparameters $\%_{val}$ and $\%_{test}$ refer to the percentage of the total number of samples ($K$) that is used repectively for validation and testing, the remaining are used for training. $N_{batch}$ is used to select the batch size and $N_{epochs}$ to select the number of epochs to use for training, i.e., the training length. When using the rolling window technique (previous Fig. \ref{fig:ML3}), it is necessary to specify the number of input samples for each window $k$ and the number of target samples (time ahead predictions) $p$. 

Both the RNN and CNN models have a feature called $Hyper$ that allows for automatic hyperparameterization of the remaining features. The $HiddenAF$ parameter sets the activation function of the hidden layers, while $OutAF$ sets the activation function of the output layer. In the case of the RNN model, the $N_{neurons}$ parameter specifies the number of neurons to use in the LSTM layer. Additionally, the $SharedDim$ hyperparameter modifies the number of neurons in the models. The $lr$ parameter specifies the learning rate, and $lf$ determines the loss function used for training the models.

For the $lf$ parameter, a new physics-aware loss function called $PA-MSE$ has been implemented specifically for reacting flows databases. This loss function aims to achieve a good reconstruction while ensuring that the sum of the species is equal to $1$. In ModelFLOWs-app, this loss function can be selected as $custom_loss$.

If the $Hyper$ parameter is set to $Yes$, these last six hyperparameters can be automated. Otherwise, they must be explicitly specified.

See more details about this application in Ref. \cite{paperAdricombustion}.

\subsubsection{Calibration of models in framework FullydDL}

\begin{table}[H]
    \centering
    \begin{tabular}{|l  l  l |}
    \hline
    \rowcolor{Gray}
    \hline
    \textbf{Parameter} & \textbf{Symbol} & \textbf{Value}
    \\ \hline \hline
        Training size               &   $\%_{train}$    &   0.8     \\
        Validation size             &   $\%_{val}$       &   0.2     \\
        Batch size                  &   $N_{batch}$     &   5      \\
        Type of model               &   $Model_{T}$     &   rnn     \\
        \# epochs                   &   $N_{epochs}$    &   140     \\
        \# input samples            &   $k$             &   10      \\
        \# time ahead predictions   &   $p$             &   2       \\
        Hyper parameterization      &   $Hyper$         &   No      \\
        Hidden activation functions (only for $Hyper$ = No) &   $HiddenAF$      &   Dense   \\
        Output activation functions (only for $Hyper$ = No) &   $OutAF$         &   None    \\
        \# Neurons (only for RNN and when $Hyper$ = No)     &   $N_{neurons}$   &   400     \\   
        Shared dimension (only for $Hyper$ = No)            &   $SharedDim$     &   80      \\
        Learning rate (only for $Hyper$ = No)               &   $lr$            &   0.005   \\
        \hline
    \end{tabular}
    \caption{Hyperparameters of models belonging to the FullyDL framework. The symbol $\#$ means number and the Value column shows the configuration used in Ref. \cite{MataLeon2023}.}
    \label{tab:calibration_FullyDL}
\end{table}

Similar to the previous subsection, in this one we summarize the hyperparameters available, listed on Tab. \ref{tab:calibration_FullyDL}, for both models, RNN and CNN, in the FullyDL framework. Some of them share the same meaning as in previous subsection, these are $N_{batch}$, $N_{epochs}$, $k$, $p$, $Hyper$, $HiddenAF$, $OutAF$, $SharedDim$ and $lr$. 
Hyperparameters $\%_{train}$ and $\%_{val}$ specify which percentage of samples  will be used for training and validation, respectively. While the entire database is used for testing. Also, in this framework the hyperparameter $Model_{T}$ specifies which model to use: RNN or CNN. Finally, $N_{neurons}$ shows the number of neurons used in the LSTM layer, inside the RNN model. Similar to HybridDL framework, the four bottom hyperparameters in Tab. \ref{tab:calibration_FullyDL} can be automated if $Hyper = \hbox{Yes}$, otherwise they must be specified.

\section{Conclusions\label{sec:conclusions}}

This article introduces the ModelFLOWs-app, a novel software tool that has proven to be effective in developing accurate and robust fully data-driven hybrid reduced order models. The software combines modal decomposition and deep learning strategies to uncover key patterns in dynamical systems, enabling a deeper understanding of the underlying physics. It offers valuable capabilities such as database reconstruction from sparse measurements and accurate prediction of system dynamics, providing an alternative to computationally expensive numerical simulations.

Although initially developed for analyzing turbulent flows, ModelFLOWs-app has demonstrated its versatility and exceptional properties in a wide range of industrial applications involving complex non-linear dynamical systems. Examples include various applications in fluid dynamics such as temporal forecasting in wind turbines, predictions in reactive flows and future projections in compressible flows with buffeting, identification of flow instabilities in turbulent flows, medical imaging pattern identification and reconstruction, wind velocity prediction using lidar measurements, identification of flutter instability in flight tests, et cetera. 

Readers are encouraged to thoroughly explore the applications presented in this article and further enhance their knowledge by accessing the tutorials and videos available on the ModelFLOWs-app website \cite{ModelFLOWsappWeb}.

\section[]{Acknowledgements}
The authors would like to acknowledge the collaboration of the following researchers, who have contributed by sharing databases, writing articles, providing new ideas, and engaging in fruitful discussions. The collective work carried out with these researchers over the past years has greatly enhanced the robustness of the current codes. These researchers are: Prof. J.M. Vega (UPM), Dr. R. Vinuesa (KTH), Prof. A. Parente (ULB), Prof. L. Brant (KTH), Dr. M. Rosti (OIST), Prof. O. Tammisola (KTH), and Prof. J. Soria (Monash Uni.).
The authors would also like to express their gratitude to the research group ModelFLOWs for their valuable discussions, assistance in generating new databases, and for their support in testing some of the developed tools. S.L.C., A.C. and S.R.A. acknowledge the grant PID2020-114173RB-I00 funded by MCIN/AEI/ 10.13039/501100011033 and the support of Comunidad de Madrid through the call Research Grants for Young Investigators from Universidad Politécnica de Madrid. A.C. also acknowledges the support of Universidad Politécnica de Madrid, under the programme ‘Programa Propio’. E.L. and S.L.C. acknowledge the support provided by Grant No. TED2021-129774B-C21 and by Grant No. PLEC2022-009235, funded by MCIN/AEI/10.13039/501100011033 and by the European Union
“NextGenerationEU”/PRTR.


\bibliographystyle{elsarticle-num-names} 
\bibliography{jfmBib}





\end{document}